\documentclass[mathleft]{an}
\usepackage{graphicx}
\usepackage{times}
\usepackage{url}
\begin{document}
 \Pagespan{005}{022} 
 \Yearpublication{2015}%
 \Yearsubmission{2014}%
 \Month{January}%
 \Volume{336}%
 \Issue{1}%

 \title{Geometric characterization of the Arjuna orbital domain
       }
 \author{C. de la Fuente Marcos\thanks{Corresponding author:\email{nbplanet@fis.ucm.es}}
         \and
         R. de la Fuente Marcos
        }
 \authorrunning{de la Fuente Marcos \& de la Fuente Marcos}
 \titlerunning{Characterization of the Arjuna orbital domain}
 \institute{Universidad Complutense de Madrid,
            Ciudad Universitaria, E-28040 Madrid, Spain
           }

 \received{2014 Jul 14}
 \accepted{2014 Oct 15}
 \publonline{2015 Feb 09}


 \abstract
   {Arjuna-type orbits are characterized by being Earth-like, having both
    low-eccentricity and low-inclination. Objects following these 
    trajectories experience repeated trappings in the 1:1 commensurability 
    with the Earth and can become temporary Trojans, horseshoe librators,
    quasi-satellites, and even transient natural satellites. Here, we review 
    what we know about this peculiar dynamical group and use a Monte Carlo 
    simulation to characterize geometrically the Arjuna orbital domain, 
    studying its visibility both from the ground and with the European Space 
    Agency Gaia spacecraft. The visibility analysis from the ground together 
    with the discovery circumstances of known objects are used as proxies to 
    estimate the current size of this population. The impact cross-section of 
    the Earth for minor bodies in this resonant group is also investigated. 
    We find that, for ground-based observations, the solar elongation at 
    perigee of nearly half of these objects is less than 90\degr. They are 
    best observed by space-borne telescopes but Gaia is not going to improve 
    significantly the current discovery rate for members of this class. Our 
    results suggest that the size of this population may have been 
    underestimated by current models. On the other hand, their intrinsically 
    low encounter velocities with the Earth induce a 10--1,000-fold increase 
    in the impact cross-section with respect to what is typical for objects 
    in the Apollo or Aten asteroid populations. We estimate that their 
    probability of capture as transient natural satellites of our planet is 
    about 8\%.
   }

 \maketitle

 \section{Introduction}
    The detection of asteroid 2013~MZ$_{5}$ on 2013 June 18 (Linder et al. 2013) marked a milestone on near-Earth object (NEO) studies as it 
    happened to be the 10\,000th NEO ever discovered.\footnote{\url{http://www.jpl.nasa.gov/news/news.php?release=2013-207}} Unfortunately, 
    less than 1\% of the 30~m-sized NEOs have been detected so far and, although unlikely to have global consequences, impacts of objects 
    this size can still cause significant devastation if they fall in populated areas (for a recent review on the subject of NEO impacts, 
    see e.g. Perna, Barucci \& Fulchignoni 2013). In the wake of the Chelyabinsk event, it is now becoming increasingly clear that our 
    planet suffers an enhanced hazard from small impactors (e.g. Brown et al. 2013; Madiedo et al. 2014; Mainzer et al. 2014). 

    Impact hazards associated to well known minor bodies can be easily evaluated but predicting the risk associated to objects that we do 
    not know is much more challenging. The most intimidating of the unknown objects are those intrinsically difficult to discover due to 
    their inherent faintness and long synodic period (e.g. Lewis 2000). The synodic period of an object relative to the Earth is the time 
    interval for the object to return to the same position as seen from our planet (e.g. Green 1985). The synodic period, $S$, is given by 
    $S^{-1} = |T^{-1} - T_{\rm E}^{-1}|$, where $T$ and $T_{\rm E}$ are the orbital periods of the object and the Earth, respectively. For a
    given minor body, it is the characteristic time-scale between favourable visibility windows. 

    The existence of a near-Earth asteroid belt made of objects with diameters under 50~m and moving in Earth-like trajectories with 
    low eccentricity was first noticed by Rabinowitz et al. (1993). Out of this relatively large population, the subset of objects moving 
    in both low-eccentricity and low-inclination orbits represents the most challenging, observationally speaking, group. Members of this 
    dynamical sub-class of small near-Earth asteroids certainly qualify as intrinsically difficult to discover because they have values of 
    the synodic period in the decades range or longer. They were unofficially termed as Arjunas---after the hero of Hindu epic poem 
    Mahabharata---by Tom Gehrels (Rabinowitz et al. 1993; Gladman, Michel \& Froeschl\'e 2000) and share nearly Earth-like orbits. The 
    dynamical evolution of this unusual group of asteroids has been explored by Tancredi (1997), Michel \& Froeschl\'e (2000), Brasser \& 
    Wiegert (2008), Kwiatkowski et al. (2009), and Granvik, Vaubaillon \& Jedicke (2012). NEOs with synodic periods in excess of 43 yr and
    moving in both low-eccentricity and low-inclination orbits can become co-orbital companions to the Earth (de la Fuente Marcos \& de la 
    Fuente Marcos 2013a), i.e. follow resonant paths.

    In this paper, we review what we know about this peculiar sub-class of near-Earth asteroids and use a Monte Carlo simulation to 
    characterize geometrically their orbital domain, studying both their visibility---from the ground and with the European Space Agency 
    (ESA) Gaia spacecraft---and membership, focusing on the transient resonant population. We also investigate the impact cross-section of 
    the Earth for objects in this group. This paper is organized as follows. Section 2 is a detailed review, aimed at placing our results 
    into context, that includes some observational data on objects moving in Arjuna-type orbits. In Sect. 3, we describe our Monte Carlo 
    scheme to compute the minimum orbit intersection distance (MOID) or perigee which is our main tool to study the visibility of 
    hypothetical objects in this group. Some closely connected issues regarding the visibility and prospective detectability of these minor 
    bodies both from the ground and with Gaia are discussed in Sects. 4 and 5; the possible size of this population is also estimated. The 
    low-velocity encounters with the Earth and the gravitational focusing factor of objects moving in Arjuna-type orbits are studied in 
    Sect. 6. A comparative analysis of the gravitationally-enhanced impact cross-section of the Earth for Arjunas, Amors, Apollos, Atens, 
    and Atiras is presented in Sect. 7. In Sect. 8 we explore the contribution of asteroids moving in Arjuna-type orbits to the overall flux 
    of Earth's impactors. Section 9 presents a discussion of our results. In Sect. 10, we summarize our conclusions.

 \section{Arjuna-type asteroids: a sub-class of near-Earth asteroids}
    There is no official asteroid class named Arjuna. The name was suggested by T. Gehrels, the director of the Spacewatch project, in 1993
    to refer to small near-Earth asteroids following nearly Earth-like orbits (Cowen 1993; Rabinowitz et al. 1993; Gladman et al. 2000). At 
    that time, J.V. Scotti, one of the Spacewatch researchers, believed that these small objects may have their origin on the surface of the 
    Moon and be the result of large asteroid impacts; in contrast, Gehrels suggested that these objects were secondary fragments of 
    asteroids originally part of the main asteroid belt, that left their formation area under the influence of Jupiter's gravity (Cowen 
    1993). Rabinowitz (1994) argued that only 5--10\% of small near-Earth asteroids follow low-eccentricity orbits. The first object 
    unofficially termed as Arjuna is 1991 VG, discovered on 1991 November 6. 

    \subsection{Arjunas in the literature: a review}
       The first minor body identified as moving in an Earth-like orbit is 1991 VG (Tancredi 1997). This Apollo asteroid follows an orbit 
       with a semi-major axis of 1.027 AU, an eccentricity of 0.049, and an inclination of 1\fdg445. Tancredi (1997) suggested that this 
       object may be a piece of Lunar ejecta. Although it is now outside Earth's co-orbital region, it may have been trapped in the 
       1:1~commensurability with our planet in the past or will become trapped in the future (de la Fuente Marcos \& de la Fuente Marcos 
       2013a). 

       Michel \& Froeschl\'e (2000) studied the origin of small near-Earth asteroids that approach our planet following low-eccentricity 
       orbits. They found that the 3:1 and $\nu_6$ resonances in the main asteroid belt, and the Mars-crosser population cannot supply 
       enough objects to explain the observed abundance; therefore, this population cannot be the result of a steady state distribution. 
       They proposed several mechanisms including collisional disruption and tidal splitting of Earth-crossers, and also observational 
       biases to account for the discrepancies observed.

       Brasser \& Wiegert (2008) studied asteroids on Earth-like orbits, also focusing on 1991 VG, and their origin but they excluded 
       resonant objects. They concluded, following Bottke et al. (1996), that low-inclination Amor- or Apollo-class objects are the most 
       likely source for this population. However, they estimated that the probability of a near-Earth asteroid ending up on an Earth-like 
       orbit is about 5$\times$10$^{-5}$. As for the impact probability, these authors found that even if low, it is much higher---up to two 
       orders of magnitude higher in some cases---than that of typical NEOs. Following Rabinowitz et al. (1993), they predicted the 
       existence of a few hundred minor bodies following Earth-like trajectories.

       Kwiatkowski et al. (2009) focused on 2006 RH$_{120}$, an asteroid that was temporarily captured into a geocentric orbit from July 
       2006 to July 2007. Their calculations indicate that this object cannot be Lunar ejecta and they proposed that its capture as 
       transient satellite could be the result of aerobraking in the Earth's atmosphere of an object previously moving in a typical 
       Earth-crossing orbit with very low MOID, a low-eccentricity Amor or Apollo minor body. 

       Ito \& Malhotra (2010) postulated the existence of a population of objects moving in highly Earth-like orbits in order to explain the
       abundance of young Lunar craters resulting from impacts caused by objects moving in very low relative velocity (with respect to the 
       Moon) trajectories. Their findings are inconsistent with widely used predictions from a model discussed in Bottke et al. (2002). 
       However, Bottke et al.'s results are not applicable to objects with absolute magnitude $>$ 22 (this critical issue is further 
       discussed in Sect. 9). Mainzer et al. (2012) have found solid observational evidence supporting Ito \& Malhotra's results. They 
       explain the excess of potentially hazardous asteroids (PHAs) moving in low-inclination orbits as resulting from the breakup of a 
       kilometre-sized body in the main asteroid belt. Additional observational evidence on the population of tiny NEOs based on NEOWISE 
       data has been recently presented by Mainzer et al. (2014). 

       Granvik et al. (2012) studied for the first time the subject of the existence of a population of irregular natural satellites of the 
       Earth. They concluded that temporarily-captured satellites are relatively common and confirmed that 2006 RH$_{120}$ remained as 
       natural satellite of our planet for about a year starting in June 2006. They also found that objects with semi-major axes very close 
       to that of the Earth are not easily capturable as transient satellites but those experiencing grazing encounters with our planet 
       (perihelion or aphelion close to the semi-major axis of the Earth) exhibit the highest capture probability. Their research concluded 
       that if there are 10$^4$ objects larger than 10 cm impacting our planet every year, nearly 0.1\% are temporarily-captured satellites 
       prior to entering the Earth's atmosphere; in contrast, low-velocity Lunar impacts for transient satellites are very unlikely. 

       The topic of objects moving in Arjuna-type orbits has been revisited by de la Fuente Marcos \& de la Fuente Marcos (2013a). Their 
       numerical simulations uncovered a previously neglected resonant family of dynamically cold small bodies. Objects in this group are 
       transient, experiencing repeated trappings in the 1:1 mean motion resonance with our planet. They can become temporary Trojans, 
       horseshoe librators, quasi-satellites, and even transient natural satellites. Periodic close encounters with the Earth--Moon system 
       trigger transitions between the various resonant states and also make this type of orbits quite unstable, with typical e-folding 
       times of 10--100 yr. This is the result of both nodes being close to the Earth's orbit. 

       In the following, we will refer to such objects as Arjuna asteroids or Arjunas; i.e. Arjuna asteroids are dynamically cold NEOs that 
       are transient---and perhaps recurrent---co-orbital companions to the Earth. However, they can also experience relatively long Kozai 
       resonance episodes (Kozai 1962) of the low-inclination type (de la Fuente Marcos \& de la Fuente Marcos 2013a). When the Kozai 
       resonance occurs at low inclinations, the argument of perihelion librates around 0\degr or 180\degr (Michel \& Thomas 1996); 
       therefore, the nodes are located at perihelion and at aphelion, i.e. away from the Earth in this case, drastically reducing the 
       probability of close encounters (see e.g. Milani et al. 1989). 

       In summary, we consider that the name Arjuna should be applied only to those asteroids experiencing recurrent resonant episodes with 
       our planet, namely the 1:1 mean motion resonance in any of its incarnations (quasi-satellite, Trojan, horseshoe librator, or hybrids 
       of them) or the low-inclination Kozai resonance (argument of perihelion librating around 0\degr or 180\degr). These objects occupy a 
       well-constrained volume in orbital parameter space (semi-major axis, $a$, eccentricity, $e$, and inclination, $i$) approximately 
       defined by: 0.985 $< a$ (AU) $<$ 1.013, 0 $< e <$ 0.1, and 0 $< i <$ 8$\fdg$56 (for further details, see de la Fuente Marcos \& de la 
       Fuente Marcos 2013a). This criterion for membership in the Arjuna-class is based on our analysis of the short-term ($\sim10^4$ yr) 
       dynamics of four relatively well-studied objects, 2003~YN$_{107}$, 2006~JY$_{26}$, 2012~FC$_{71}$, and 2013~BS$_{45}$. The orbits of 
       these objects are all sufficiently well constrained. Therefore, the criterion for membership is ultimately based on dynamical 
       calculations, i.e. theory; it is not observational and it does not necessarily reflect the currently observed density of objects 
       within that volume of the orbital parameter space. 

       The boundaries of the volume pointed out above are linked to the actual values of the osculating orbital elements of the Earth, which 
       oscillate on a time-scale of 10$^5$ yr (see e.g. Fig. 1 in de la Fuente Marcos \& de la Fuente Marcos 2012). All these objects are 
       also members of the Amor, Apollo, Aten, or Inner Earth Object (IEO, also known as Apoheles or Atiras) populations (see Sect. 7 for 
       details on their respective orbital elements domains); objects in the Amor or IEO classes do not cross the orbit of our planet. As a 
       result of the near commensurability, the synodic periods of these minor bodies exceed 43~yr. Some of them can easily evade detection 
       for thousands of years and never pass close enough to the Earth for discovery before they hit. For the Arjunas, and due to their long 
       synodic periods, it is very important to perform extensive observations during the discovery apparition, otherwise they can be easily 
       lost. 

    \subsection{Observed objects}
       Out of the objects pointed out above, 1991~VG is currently outside the Arjuna orbital parameter domain as proposed by de la Fuente 
       Marcos \& de la Fuente Marcos (2013a) but 2006~RH$_{120}$ is well inside (see Table \ref{members}). As of 2014 July 8, there are 13 
       known objects within the boundaries of the orbital parameter space described above (see Table \ref{members}). All these objects 
       follow paths that are crossing the orbit of the Earth, i.e. there are no Amors or Atiras according to the Jet Propulsion Laboratory 
       (JPL) Small-Body Database\footnote{\url{http://ssd.jpl.nasa.gov/sbdb_query.cgi}}, and they are almost evenly distributed between the 
       Apollo and Aten dynamical classes (see Table \ref{discovery}). However, these objects can easily switch between dynamical classes 
       during close encounters with the Earth--Moon system; for example, an Apollo can become an Amor or an Aten turn into an Apollo after a 
       close flyby. The smallest predicted MOID with respect to our planet, just 0.00005 AU, is found for 2006 JY$_{26}$. Asteroid 
       2013~RZ$_{53}$ has the longest synodic period, over a thousand years. The largest known are 2014~EK$_{24}$, 2009~SH$_{2}$, 
       2012~FC$_{71}$, 2013~BS$_{45}$, 2010~HW$_{20}$, 2012~LA$_{11}$, and 2003~YN$_{107}$; all of them have most probable sizes in the 
       range 10--60~m, the only exception is the recently discovered Aten asteroid 2014~EK$_{24}$ that may be larger, 60--150~m. These 
       objects are big enough to cause extensive local damage if they fall. 
       
%
%
       \begin{table*}
        \centering
        \fontsize{8}{11pt}\selectfont
        \tabcolsep 0.30truecm
        \caption{Orbital properties of objects moving in Arjuna-type orbits (source: JPL Small-Body Database, source for $\Delta$-$v$: 
                 {\scriptsize http://echo.jpl.nasa.gov/$\sim$lance/delta\_v/delta\_v.rendezvous.html}). The orbital elements have been
                 computed at Epoch 2456800.5 (2014-May-23.0) with the exception of 2006 RH$_{120}$ (2454115.5) and 2009 BD (2455200.5).
                }
        \begin{tabular}{cccccccc}
         \hline
           Object          & $a$ (AU) &   $e$    & $i$ (\degr) & $\Omega$ (\degr) & $\omega$ (\degr) & MOID (AU) & $\Delta$-$v$ (km s$^{-1}$) \\
         \hline
           2003 YN$_{107}$ & 0.988697 & 0.013988 &   4.32115   &     264.42752    &    87.48882      & 0.004436  &  4.879                     \\
           2006 JY$_{26}$  & 1.010021 & 0.083042 &   1.43927   &      43.47229    &   273.60220      & 0.000052  &  4.364                     \\
           2006 RH$_{120}$ & 0.998625 & 0.019833 &   1.52613   &     290.52215    &   177.87923      & 0.000679  &  3.820                     \\
           2008 KT         & 1.010800 & 0.084807 &   1.98425   &     240.63099    &   102.10156      & 0.000705  &  4.425                     \\
           2008 UC$_{202}$ & 1.010179 & 0.068692 &   7.45350   &      37.36397    &    91.66396      & 0.002890  &  5.471                     \\
           2009 BD         & 1.008614 & 0.040818 &   0.38516   &      58.48799    &   110.50392      & 0.003565  &  3.870                     \\
           2009 SH$_{2}$   & 0.991719 & 0.094175 &   6.81073   &       6.70423    &   101.56545      & 0.000390  &  5.070                     \\
           2010 HW$_{20}$  & 1.010924 & 0.050111 &   8.18503   &      39.23805    &    60.24051      & 0.008984  &  5.690                     \\
           2012 FC$_{71}$  & 0.989158 & 0.088009 &   4.94929   &      38.33286    &   347.58558      & 0.056850  &  4.686                     \\
           2012 LA$_{11}$  & 0.987936 & 0.096338 &   5.10630   &     260.94868    &   241.57485      & 0.007504  &  4.751                     \\
           2013 BS$_{45}$  & 0.993858 & 0.083801 &   0.77333   &      83.55476    &   149.77660      & 0.011514  &  4.083                     \\
           2013 RZ$_{53}$  & 0.999722 & 0.048260 &   1.50653   &     357.44138    &    89.18110      & 0.001895  &  4.198                     \\
           2014 EK$_{24}$  & 1.003690 & 0.072150 &   4.72541   &     342.05892    &    62.67750      & 0.033854  &  4.846                     \\
         \hline
        \end{tabular}
        \label{members}
       \end{table*}
%
%
%
%
       \begin{table*}
        \centering
        \fontsize{8}{11pt}\selectfont
        \tabcolsep 0.30truecm
        \caption{Equatorial coordinates, apparent magnitudes (with filter), solar elongation, $\theta$, and phase, $\phi$, at discovery 
                 time ($t_d$) for all the known objects currently moving in Arjuna-type orbits. The dynamical class assigned by the JPL
                 Small-Body Database, the synodic period, $S$, and the absolute magnitude, $H$, are also indicated (J2000.0 ecliptic and 
                 equinox. Sources: Minor Planet Center (MPC) and JPL Small-Body Databases).
                }
        \begin{tabular}{cccccccccc}
         \hline
           Object          & $\alpha$ ($^{\rm h}$:$^{\rm m}$:$^{\rm s}$) & $\delta$ (\degr:\arcmin:\arcsec) & $t_d$      &$m$ (mag) & $\theta$(\degr) & $\phi$(\degr) 
                           & Class  & $S$ (yr) & $H$ (mag) \\
         \hline
           2003~YN$_{107}$ & 11:04:54.44                                 & +10:24:31.0                      & 2003-12-20 & 18.8      & 104.7           &  74.4
                           & Aten   &   56.767 & 26.5      \\
           2006~JY$_{26}$  & 14:17:08.91                                 & -06:47:25.2                      & 2006-05-06 & 18.0 (V)  & 166.9           &  13.0
                           & Apollo &   69.315 & 28.4      \\
           2006~RH$_{120}$ & 23:11:31.09                                 & -07:31:12.5                      & 2006-09-14 & 19.3 (V)  &  73.4           & 106.4
                           & Aten   &  508.787 & 29.5      \\
           2008~KT         & 15:32:01.68                                 & -10:21:18.7                      & 2008-05-28 & 19.7 (V)  & 163.8           &  16.0
                           & Apollo &   64.241 & 28.2      \\
           2008~UC$_{202}$ & 02:32:14.36                                 & +10:36:56.9                      & 2008-10-31 & 20.1      & 175.7           &   4.3
                           & Apollo &   68.219 & 28.3      \\
           2009~BD         & 07:53:19.03                                 & +21:35:31.9                      & 2009-01-16 & 18.4 (V)  & 179.3           &   0.6
                           & Apollo &  108.083 & 28.1      \\
           2009~SH$_{2}$   & 03:13:18.76                                 & -32:57:05.2                      & 2009-09-18 & 18.8 (V)  & 122.0           &  56.5
                           & Aten   &   77.107 & 24.9      \\
           2010~HW$_{20}$  & 16:11:25.36                                 & +01:59:59.2                      & 2010-04-25 & 20.2 (V)  & 146.6           &  32.6
                           & Apollo &   63.506 & 26.1      \\
           2012~FC$_{71}$  & 14:42:45.32                                 & +13:21:13.9                      & 2012-03-31 & 21.8 (V)  & 144.8           &  32.5
                           & Aten   &   59.158 & 25.2      \\
           2012~LA$_{11}$  & 15:33:16.84                                 & -03:50:09.1                      & 2012-06-15 & 18.3      & 144.7           &  34.8
                           & Aten   &   53.216 & 26.1      \\
           2013~BS$_{45}$  & 10:26:27.30                                 & +19:20:12.2                      & 2013-01-20 & 20.4 (V)  & 148.0           &  30.7
                           & Aten   &  103.073 & 25.9      \\
           2013~RZ$_{53}$  & 23:44:39.23                                 & -03:33:17.0                      & 2013-09-13 & 20.2 (V)  & 175.2           &   4.8
                           & Aten   & 1197.030 & 31.1      \\
           2014~EK$_{24}$  & 12:59:36.58                                 & -20:37:14.3                      & 2014-03-10 & 17.9 (V)  & 145.4           &  32.7
                           & Apollo &  196.430 & 23.2      \\
         \hline
        \end{tabular}
        \label{discovery}
       \end{table*}
%
%

\begin{figure}
          \resizebox{\hsize}{!}{
              \includegraphics{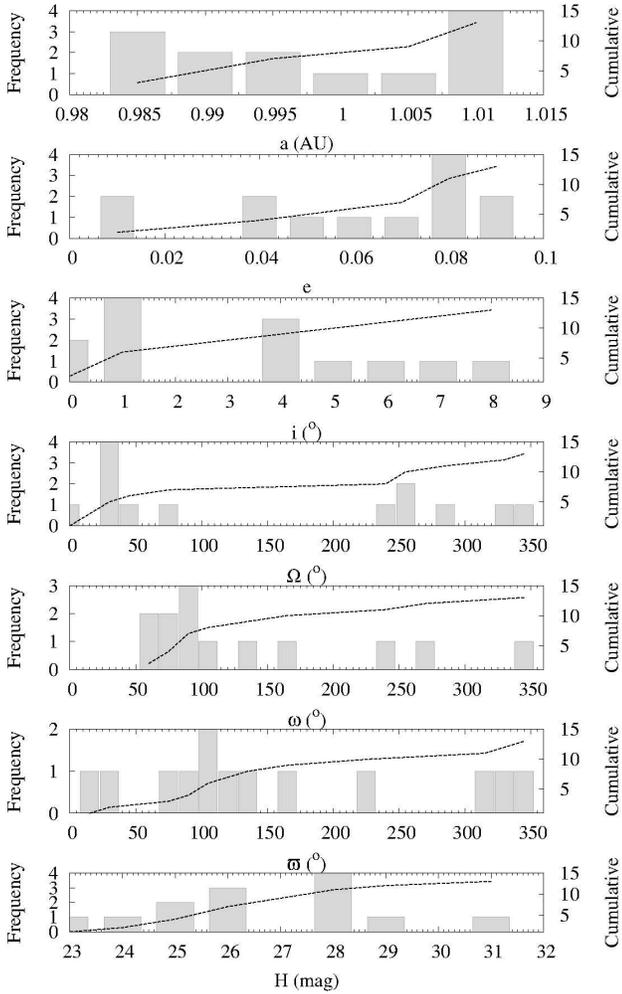}
          }
          \caption{Distributions in semi-major axis, $a$, eccentricity, $e$, inclination, $i$, longitude of the ascending node, $\Omega$,
                   argument of perihelion, $\omega$, longitude of perihelion, $\varpi$, and absolute magnitude, $H$, of the objects in Table 
                   \ref{members}.
                  }
          \label{histo}
       \end{figure}

       Figure \ref{histo} shows the distribution in orbital parameter space and absolute magnitude, $H$, of the objects in Table 
       \ref{members}. Even if the number of known objects is small, some trends are visible. The semi-major axes are not uniformly 
       distributed (most objects are near the edges of the orbital domain), approximately half (6/13) the objects have eccentricities near 
       the maximum value and also nearly half (6/13) the objects have inclinations under 2\degr. As for both the longitude of the ascending 
       node and the argument of perihelion, the distributions are far from uniform which is consistent with recent results obtained by
       JeongAhn \& Malhotra (2014) for the overall NEO population. In particular, eight objects have argument of perihelion close to 
       90\degr. This means that these objects reach perihelion at their northernmost distance from the ecliptic. Two other objects have 
       argument of perihelion close to 270\degr, i.e. their perihelia happen southernmost from the ecliptic. In other words, 10 objects out 
       of 13 reach perihelion far from the nodes. This is at odds with the overall distribution of arguments of perihelion of observed NEOs 
       found by JeongAhn \& Malhotra (2014), indicating that Arjunas are a somewhat distinct dynamical group. The values of the longitude of 
       the ascending node of four objects are close to that of the Earth. The distribution in absolute magnitude, which is a proxy for size, 
       clearly shows a regular pattern. 
%
%
%
%

       The objects in Table \ref{members} have not been discovered by a single survey and their discovery is not the result of an 
       observational strategy specifically aimed at detecting minor bodies moving in both low-eccentricity and low-inclination, Earth-like 
       orbits. They have been found as a by-product of searching for objects that might cross the Earth's path. It is very difficult to 
       understand how the complex and heterogeneous pointing history of the five different surveys (Mt. Lemmon Survey, Catalina Sky Survey, 
       Lincoln Laboratory Experimental Test System in New Mexico, Spacewatch at the Steward Observatory in Kitt Peak, and Siding Spring 
       Survey) involved in the discovery of the asteroids in Table \ref{members} maps the three-dimensional ($a$, $e$, and $i$) orbital 
       elements distribution into actual biases for these three orbital elements. 

       As for the longitude of the ascending node and the argument of perihelion, seasonal biases affecting the distribution of these 
       elements for the general class of NEOs have been extensively discussed by JeongAhn \& Malhotra (2014). They have found that nearly 
       1/3 of NEOs were discovered between September and November. Over 1/3 have been discovered between December and March and just 14\% 
       between June and August. These numbers match well those in Table \ref{discovery}. As expected, observational biases associated to 
       the Arjuna orbital parameter space are the same as those affecting the overall NEO population. However, JeongAhn \& Malhotra (2014) 
       find peaks for $\Omega$ in the interval 170--220\degr and 0--40\degr. Figure \ref{histo} shows no objects or, comparatively, very few 
       in those ranges (see Fig. 4 in JeongAhn \& Malhotra 2014). In sharp contrast, most observed Arjunas appear clustered towards values 
       of $\Omega$ where a lower fraction of objects is found by JeongAhn \& Malhotra (2014). That distribution is tentatively explained by
       JeongAhn \& Malhotra as the result of secular perturbations by Jupiter. Within the general NEO population, the Arjunas may be 
       comparatively less affected by Jupiter; in the case of Kozai librators, Jupiter must be the source of the stabilising force. 

       The distribution in argument of perihelion in Fig. \ref{histo} does not match that of faint NEOs in Fig. 11 in JeongAhn \& Malhotra 
       (2014) well. The peak at 90\degr is observed but no cluster around 0\degr, 180\degr, or 270\degr is present. Even if based on small 
       number statistics, it seems clear that the distribution observed in Fig. \ref{histo} can not have its origin in the observational 
       bias due to their orbital geometry as described in JeongAhn \& Malhotra (2014). It must obey to some other effect or perhaps be 
       intrinsic to this population. Consistently, a similar behaviour is observed for the longitude of perihelion, $\varpi = \Omega + 
       \omega$ (compare Fig. 17 in JeongAhn \& Malhotra 2014 and the second to last panel in Fig. \ref{histo}).

       The distributions of the orbital elements of the known Arjunas are strongly non-uniform and those of the angular elements ($\Omega$,
       $\omega$, and $\varpi$) appear to be incompatible with the ones of the general group of known NEOs. For that reason, it is unlikely 
       that correcting the Arjuna distributions using the same approach as in the case of general NEOs may help our analysis. On the other
       hand, some of the objects currently moving within the Arjuna orbital domain appear to be fragments of fragments, perhaps generated
       within the neighbourhood of the Earth--Moon system (see Sect. 9), and this origin makes the application of any debiasing technique
       intrinsically difficult as they do not include the possibility of secondary fragmentation. Correcting the distributions in Fig. 
       \ref{histo} by debiasing the sample population in Table \ref{members} is certainly beyond the scope of this work and it will not be 
       attempted here. 

       Instead of using a debiasing approach as described in, e.g., Jedicke \& Metcalfe (1998), Spahr (1998), Bottke et al. (2002), Mainzer 
       et al. (2011) or Grav et al. (2011), we discuss the effects of the most obvious sources of bias (seasonal variation of discovery 
       rates, relative geometry, uneven distribution of telescopes on the surface of our planet, magnitude, or synodic period) in Sect. 4 
       and elsewhere. It must be noticed that the distribution functions found in those studies (including Bottke et al. 2002) assume that 
       the angular elements are randomly and uniformly distributed in the interval (0, $2\pi$) which is clearly incorrect (JeongAhn \& 
       Malhotra 2014). In the following sections, a discussion of the effects of survey biases is presented each time that our geometrical 
       results are compared to the observational ones.

       On a more practical side, Arjuna-type asteroids are among the best possible targets in terms of propellant budget for asteroid 
       exploration missions. In general, and for any celestial body, the suitability for spacecraft visitation is characterized by the 
       delta-$v$ ($\Delta$-$v$) scalar parameter that measures the total change in velocity required to complete the mission (Shoemaker \& 
       Helin 1978). This parameter is independent of the mass of the spacecraft. The delta-$v$ required to reach the most favourable of 
       these objects is nearly 3.8 km s$^{-1}$, e.g. 2006 RH$_{120}$ and 2009 BD (see Table \ref{members}); in contrast, the delta-$v$ 
       necessary to transfer from a low-Earth orbit to the Moon is 6.0 km s$^{-1}$. Asteroid 2006 RH$_{120}$ is clearly of natural origin, 
       not human-made orbital debris (Kwiatkowski et al. 2009); the same can be said about 2009 BD (Micheli, Tholen \& Elliott 2012; Mommert 
       et al. 2014). However, the long synodic periods (see Table \ref{discovery}) place severe constraints for the design and scheduling of 
       rendezvous missions to these objects. The same limitations apply to any hypothetical asteroid redirect mission aimed at deflecting an 
       incoming object moving in an Arjuna-type orbit.

       Recapitulating, Arjuna-type asteroids are interesting as relatively easy targets for space exploration missions and for being 
       observationally challenging, potentially hazardous objects. At present, both the dynamical origin and the current size of this 
       peculiar population are controversial issues, far from well established. It is often said that widely accepted models (e.g. Bottke et 
       al. 2002) predict a tiny size for this population; however, recent observational evidence appears to be inconsistent with that idea. 
       In any case, Bottke et al. (2002) predictions are not applicable to objects with absolute magnitude $>$ 22 (see Sect. 9) as it is the 
       case of all the objects in Table \ref{members}. Dynamical origin and membership (size of the population) are likely related issues. 
       They cannot be primordial because they follow rather chaotic paths with very short e-folding times (10--100 yr); as an intrinsically 
       transient population, one or more mechanisms should be at work to mitigate the losses, repopulating the group.  

 \section{Computing the perigee: a Monte Carlo approach}
    Arjuna asteroids have orbits very similar to Earth's and they are all small, with most probable sizes in the range 10--60~m. Therefore, 
    they can only be properly observed at perigee. Any analysis of the visibility of these objects demands the prior study of their 
    perigees. In other words, the first question to be answered is that of, given a set of orbits (constrained in semi-major axis, 
    eccentricity, and inclination), where on the celestial sphere are they reaching perigee? In our case, the orbital parameter domain is 
    constrained by the criterion described in de la Fuente Marcos \& de la Fuente Marcos (2013a) which is of theoretical nature, not the 
    result of an observational survey in search for objects moving in Arjuna-type orbits (see above). Calculating the perigee of a given 
    orbit is formally equivalent to computing the MOID of the object with respect to the Earth. The answer to this purely geometrical 
    question is, in principle, independent from the distributions in Fig. \ref{histo}.

    In general, computing the MOID is far from trivial and it amounts to finding the minimum separation between two Keplerian orbits in 
    three-dimensional space (see, e.g., Kholshevnikov \& Vassiliev 1999; Gronchi \& Valsecchi 2013a,b). The solution of this problem does
    not have a closed form but it is well suited for a Monte Carlo simulation (Metropolis \& Ulam 1949; Press et al. 2007) using the 
    equations of the orbit of an object around the Sun under the two-body approximation. If these equations (for both the Earth and the 
    object) are extensively sampled in phase space, the usual Euclidean distance between any two points on the orbits can be systematically 
    calculated until the minimal distance (MOID or perigee in this case) is eventually found. Using the perigee, both the visibility and the 
    impact cross-section of the Earth for these objects can be studied. Comparing the distribution of perigees (objects in this group are so 
    small that they can only be detected at perigee) and orbital elements with those of real, observed objects (see Tables \ref{members} and 
    \ref{discovery}, and Fig. \ref{histo}), makes the identification of biases and truly intrinsic properties of this population easier. 

    Under the two-body approximation, the motion of an object around the Sun in space (the Keplerian orbit) can be described by the 
    expressions (e.g. Murray \& Dermott 1999):
    \begin{eqnarray}
       X & = & r \ \left[\cos \Omega \cos(\omega + f) - \sin \Omega \sin(\omega + f) \cos i\right]
                   \nonumber \\
       Y & = & r \ \left[\sin \Omega \cos(\omega + f) + \cos \Omega \sin(\omega + f) \cos i\right]
                   \label{orbit} \\
       Z & = & r \ \sin(\omega + f) \sin i \nonumber
    \end{eqnarray}
    where $r = a (1 - e^{2})/(1 + e \cos f)$, $a$ is the semi-major axis, $e$ is the eccentricity, $i$ is the inclination, $\Omega$ is the 
    longitude of the ascending node, $\omega$ is the argument of perihelion, and $f$ is the true anomaly. Given two sets of orbital elements 
    ($a$, $e$, $i$, $\Omega$, and $\omega$), these equations can be applied to compute the MOID. In our case, the first object is the Earth; 
    the second one being the minor body moving in an Arjuna-type trajectory. 

    Our method to compute the MOID is far more time consuming than other available algorithms but makes no a priori assumptions and can be
    applied to arbitrary pairs of heliocentric orbits. It produces results that are consistent with those from other methods. Numerical 
    routines to compute the MOID have been developed by Baluev \& Kholshevnikov (2005), Gronchi (2005), \v{S}egan et al. (2011) and
    Wi\'sniowski \& Rickman (2013), among others. Gronchi's approach is widely regarded as the de facto standard for MOID computations 
    (Wi\'sniowski \& Rickman 2013).

    The orbit of the Earth used to perform the calculations whose results are discussed here was computed at Epoch JD 2456600.5 that 
    corresponds to 0:00 UT on 2013 November 4 (Heliocentric ecliptic orbital elements, J2000.0 ecliptic and equinox, see Table \ref{Earth}) 
    and it was provided by the JPL HORIZONS system (Giorgini et al. 1996; Standish 1998).\footnote{\url{http://ssd.jpl.nasa.gov/horizons.cgi}} 
    Results obtained for epochs within a few hundred years from the one used here are virtually the same; however, for distant epochs into 
    the past or the future results are rather different due to secular changes in the orbital elements of the Earth. The osculating orbital 
    elements ($a$, $e$, $i$) of the Earth oscillate on a time-scale of 10$^5$ yr (see, e.g., Fig. 1 in de la Fuente Marcos \& de la Fuente 
    Marcos 2012). 

    For the object moving in an Arjuna-type orbit, the orbital elements are randomly sampled within fixed (assumed as described in Sect. 2 
    and de la Fuente Marcos \& de la Fuente Marcos 2013a) ranges following a uniform distribution. Then, and for each set of orbital 
    elements, we randomly sample the above equations in true anomaly (0--360\degr) for both the object and the Earth, computing the usual 
    Euclidean distance between both points so the minimal distance (in this case the perigee) is eventually found. Two different resolutions 
    were used in this Monte Carlo simulation: 10$^{5}$ and 10$^{7}$ points per orbit, which are equivalent to a time resolution of 5.26 
    minutes and 3.16 seconds, respectively. Results are consistent within the precision limits of this research. 

    A uniform distribution is used because the actual spread in orbital parameter space for these objects is not well established. Mainzer 
    et al. (2011), Greenstreet, Ngo \& Gladman (2012), or Greenstreet \& Gladman (2013) show distributions that are consistent with Bottke 
    et al.'s (2002) predictions for the type of orbits discussed here but it is unclear how the issues recently pointed out by Mainzer et 
    al. (2014) regarding the bias in favour of small objects with high albedo may affect their conclusions. Also, the results obtained by 
    JeongAhn \& Malhotra (2014) make it difficult to decide which orbital elements distributions are the best ones to investigate the 
    distribution of perigees on the celestial sphere of the peculiar orbits studied here. Even if our choice may not be very realistic, 
    uniform distributions are useful when making geometrical characterizations like the one attempted here because they can help to uncover 
    observational biases within observational data (in our case, those plotted in Fig. \ref{histo}) and also intrinsic dynamical signatures 
    associated to the studied populations. We have recently used this approach to study the visibility of objects moving in Venus' (de la 
    Fuente Marcos \& de la Fuente Marcos 2014a) and Uranus' (de la Fuente Marcos \& de la Fuente Marcos 2014b) co-orbital regions.
%
%
     \begin{table*}
      \centering
      \fontsize{8}{11pt}\selectfont
      \tabcolsep 0.1truecm
      \caption{Orbital elements of the Earth on JD 2456600.5 = A.D. 2013-Nov-4 00:00:00.0 UT (Source: JPL HORIZONS system).}
      \begin{tabular}{ccccc}
       \hline
          $a$ (AU)          & $e$                 & $i$ (\degr)          & $\Omega$ (\degr)  & $\omega$ (\degr)  \\
       \hline
          1.000934902712727 & 0.01719105606083375 & 0.003728586449775013 & 202.0300077899986 & 258.4106714905184 \\
       \hline
      \end{tabular}
      \label{Earth}
     \end{table*}
%
%

    Our results are only approximate but statistically robust and reliable enough for planning actual observations aimed at discovering more 
    objects moving in Arjuna-type orbits. The use of the two-body approximation implies that both objects (the Earth and the minor body in 
    our case) are point masses, and that external and internal forces acting upon the bodies are neglected. At this point, it can be argued 
    that using the two-body approximation for objects characterized by low (relative to the Earth) orbital velocities and short perigees is 
    entirely inappropriate because three-body effects (those arising from the gravitational interaction between the Earth--Moon system and 
    the minor body) are fundamental in shaping the relative dynamics. These minor bodies penetrate deep inside the region where that 
    approximation is no longer valid because the gravitational field of the Earth, not the Sun, is dominant. However, if we focus on the 
    region exterior to a sphere centred on the Earth and of radius equal to the Hill radius of the Earth ($r_{\rm H}$ = 0.0098 AU) which is 
    the conventional limit for the sphere of influence of our planet, that concern can be avoided. 

    Nevertheless, and for a minor body following a trajectory passing the immediate neighbourhood of our planet, the largest orbital changes 
    take place when it is within 10 Earth radii ($R_{\rm E}$ = 6,371 km) from the Earth's centre (see e.g. Jenniskens et al. 2009; 
    Oszkiewicz et al. 2012). If the predicted perigee of the minor body is outside 0.000425 AU (10\ $R_{\rm E}$), the impact of the flyby 
    with the Earth on the actual values of its orbital elements will be minor (only two objects in Table \ref{members} have MOIDs under 
    10\ $R_{\rm E}$). In any case, and if during the interaction these values change significantly, we will have one of two situations. 
    Either the new trajectory is also an Arjuna-type orbit and, therefore, it is already included in the pool of randomly generated sets of 
    orbital elements at study or the new orbit is not Earth-like and, as a result, no longer of interest here. In summary, our method is 
    clearly valid in statistical terms because it does not focus on the detailed dynamical evolution of objects moving in Arjuna-type orbits 
    but on their probabilistic, averaged properties. 

    Minor bodies moving in Arjuna-type orbits can only experience close encounters with the Earth--Moon system. Our approach makes very few, 
    if any, a priori assumptions and it does not depend on how precise our integrations are during the close encounters because it is not an 
    $N$-body simulation. For the same reasons, the effects of non-gravitational forces like the Yarkovsky force, do not have any influence 
    on our results because if an object moving in an Arjuna-type orbit is affected by the Yarkovsky force and/or radiation pressure (i.e.
    non-gravitational forces), the values of its orbital parameters will likely change smoothly to those of another, usually close, 
    Arjuna-type orbit.  

    It may also be argued that the actual minimal approach distance of a given minor body does not coincide with the MOID. In general, when 
    a close approach between a planet and a minor body takes place, it does not occur at the MOID. The MOID is the minimum possible distance 
    between the orbit of the small body and that of the planet if we neglect gravitational focusing. The unperturbed close approach distance 
    cannot be less than the MOID but for encounters under 10\ $R_{\rm E}$, and depending on the geometry of the flyby, the actual physical 
    distance between minor body and planet could become significantly smaller than the nominal MOID. Most flybys are characterized by values 
    of the distance of closest approach larger than the nominal MOID. However, we are interested in objects that can be detected. They must 
    be close to the Earth in order to be observed because they are small; therefore, only the ones undergoing flybys close to the MOID are 
    relevant for our study. 

    Estimating the actual physical distance between a minor body and a planet requires full $N$-body calculations and, therefore, many more 
    computational resources than the ones used here. $N$-body calculations are more precise but they have inherent limitations that make 
    them ill-suited for the type of statistical analysis completed in this study. Our results, being geometrical, do not depend on the size 
    of the objects moving in Arjuna-type orbits and they are equally applicable to metric-sized minor bodies, the Arjuna asteroids, or dust 
    particles in co-orbital motion with our planet.  
 
 \section{Finding them: visibility analysis for ground-based observatories}    
    The characteristics, strengths, and limitations of our Monte Carlo technique have been highlighted in the previous section. Being able 
    to compute the perigees of hypothetical objects moving in Arjuna-type orbits as described in Sect. 2.1, or de la Fuente Marcos \& de la 
    Fuente Marcos (2013b), gives us the ability to study their theoretical visibility. 

    As pointed out above, a very dangerous property of those objects moving in Arjuna-type orbits lies in their long (in excess of 43 yr) 
    synodic periods with respect to the Earth. This feature has a major impact on their visibility and eventual detectability when 
    observing from our planet. During most of the time between flybys, such objects reside in the daytime sky as seen from the Earth and, 
    therefore, they are optically unobservable from the ground. It is theoretically possible that putative, undiscovered impactors may reach 
    perigee right after coming out of the daytime sky and therefore they may hit with little or no warning; think about the Chelyabinsk 
    superbolide even if its parent body was not following an Earth-like orbit, see e.g. de la Fuente Marcos \& de la Fuente Marcos (2013b, 
    2014c). 

    Focusing on objects of meteoroid or asteroidal-size, not interplanetary dust, we now consider typical limitations found in current 
    surveys. NEOs are rather small objects and in order to maximize the detection capabilities of the available instrumentation, surveys 
    usually focus on the area of sky within 45\degr of the anti-Sun point. In some cases, this constraint is relaxed to 60\degr but rarely 
    towards the dawn or dusk sky ($\sim$90\degr from the Sun). In general, observatories (including the Hubble Space Telescope) never point 
    their telescopes within 20\degr from the Sun and rarely within 45--50\degr (this is done sometimes for comets, some asteroids, Mercury, 
    and Venus, when the scientific justification is compelling) to prevent damage to the equipment. The areas of sky within a solar 
    elongation (apparent angular separation between the object and the Sun as seen from the Earth) of 80\degr are most likely outside 
    current surveys. Also, southern declinations below -30\degr and northern declinations above +80\degr are outside the coverage of most 
    surveys. These technical limitations have an obvious impact on the discovery parameters of Arjunas (see Table \ref{discovery}): most 
    objects have been found close to the anti-Sun point and only one object was discovered at solar elongation $<$ 90\degr. On the other 
    hand, objects with argument of perihelion around 270\degr are under-represented with respect to those with argument of perihelion near 
    90\degr because they reach perigee at southern declinations below -30\degr; there are more telescopes in the northern hemisphere. 

\begin{figure}
        \resizebox{\hsize}{!}{
            \includegraphics{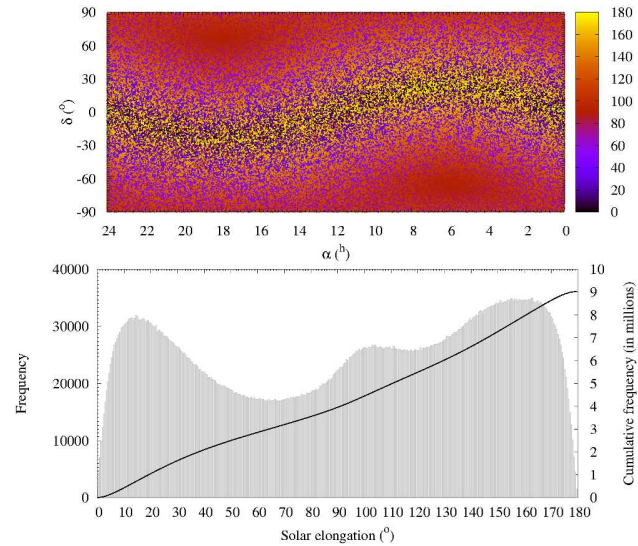}
        }
        \caption{Distribution in equatorial coordinates at perigee of Arjuna-type orbits as a function of the geocentric solar elongation 
                 (top panel). A frequency analysis of the same data (bottom panel), the bin size is 0\fdg5. Only perigees $<$ 0.05 AU are 
                 considered. Given the size of the samples studied and the values of the frequencies, error bars are too small to be seen.
                }
      \label{elongation}
    \end{figure}

\begin{figure}
        \resizebox{\hsize}{!}{
            \includegraphics{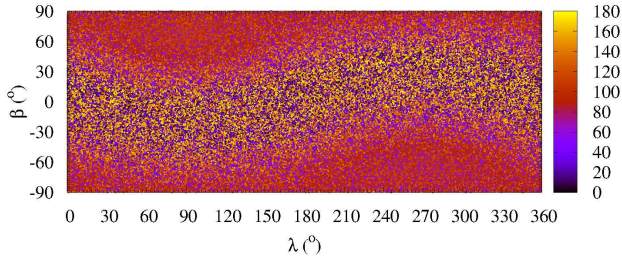}
        }
        \caption{Distribution in geocentric ecliptic coordinates at perigee of Arjuna-type orbits as a function of the geocentric solar 
                 elongation (ecliptic longitude, $\lambda$, ecliptic latitude, $\beta$).  
                }
      \label{elongationC}
    \end{figure}

    But, what is the actual influence of all these limitations on our theoretical ability to detect objects moving in Arjuna-type orbits 
    near perigee? How large is the fraction of these objects currently escaping detection? Our Monte Carlo simulation can provide reasonably 
    reliable answers to these important questions. Figure \ref{elongation} (also Fig. \ref{elongationC}) shows the results of the analysis 
    of 10 million Arjuna-like orbits uniformly distributed in orbital parameter space as described in Sect. 3; only perigees $<$ 0.05 AU are 
    considered. The distribution in geocentric equatorial coordinates at perigee as a function of the geocentric solar elongation is 
    displayed on the top panel and it shows that a significant fraction of relevant objects reach perigee within 20\degr from the Sun. A 
    larger group reaches perigee within 90\degr from the Sun and near the ecliptic poles. A frequency analysis of the same data shows actual 
    fractions. Nearly 15\% of these objects reach perigee within 20\degr from the Sun, about 50\% have perigees within 90\degr from the Sun, 
    and only 25\% of the objects reach perigee within 45\degr of the anti-Sun point. In addition, nearly 30\% of the objects reach perigee 
    with declination below -30\degr or above +80\degr. In summary, and under the assumption of uniformly distributed orbital elements, it is 
    quite possible that more than 65\% of all the relatively bright objects moving in Arjuna-like trajectories can escape detection by the 
    currently available telescopic facilities. This estimate assumes that all the relatively bright objects reaching perigee within 90\degr 
    from the anti-Sun point are detected ($\sim$50\%) but the fraction of objects reaching perigee with declination below -30\degr or above 
    +80\degr ($\sim$30\%). Because we do not know the true distribution in size and albedo, this is probably a lower limit, i.e. more than 
    65\% of all Arjunas may escape detection.

    If we compare with data in Table \ref{discovery} and assuming that all the currently known objects were discovered near perigee, we 
    observe that one object out of 13 ($\sim$8\%) reached perigee within 90\degr from the Sun and 10 out of 13 ($\sim$77\%) objects reached 
    perigee within 45\degr of the anti-Sun point. Only one object ($\sim$8\%) was found with declination $<$ -30\degr. Although the number 
    of known objects is small, it seems that our overall 35\% estimate for the current detection effectiveness for (relatively bright) 
    objects moving in Arjuna-type orbits is probably conservative; the actual efficiency is likely lower. 
%
%
%
%
%
%
%
%
%
%
    \begin{figure}
        \resizebox{\hsize}{!}{
            \includegraphics{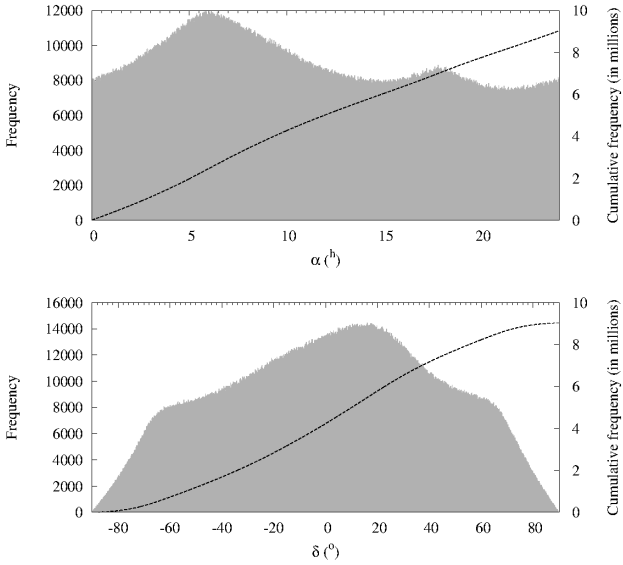}
        }
        \caption{Frequency distribution in equatorial coordinates (right ascension, top panel, and declination, bottom panel) of Arjuna-type 
                 orbits at perigee. The bin sizes are 0.024 hours in right ascension and 0\fdg18 in declination, error bars are too small to 
                 be seen. The best areas to search for these objects are located around right ascension 6$^{\rm h}$ and 18$^{\rm h}$ and 
                 declination $\in$(0, 30)\degr.
                }
      \label{radec}
    \end{figure}
%
%
%
%
    \begin{figure}
        \resizebox{\hsize}{!}{
            \includegraphics{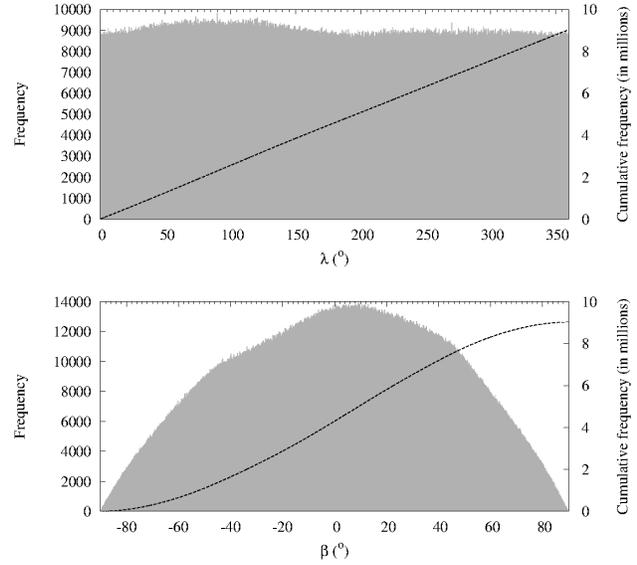}
        }
        \caption{Frequency distribution in geocentric ecliptic coordinates (longitude, top panel, and latitude, bottom panel) of Arjuna-type 
                 orbits at perigee. The bin sizes are 0\fdg36 in $\lambda$ and 0\fdg18 in $\beta$, error bars are too small to be seen. 
                }
      \label{lonlat}
    \end{figure}
%
%

\begin{figure}
        \resizebox{\hsize}{!}{
            \includegraphics{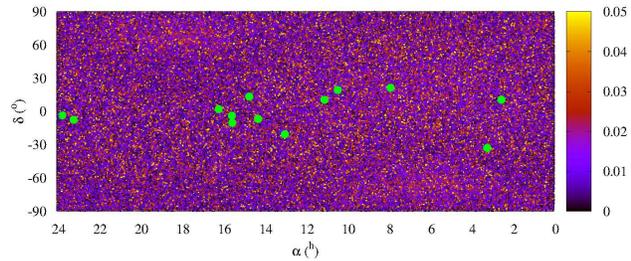}
        }
        \caption{Distribution in equatorial coordinates of the perigees (as in Fig. \ref{map}) of Arjuna-type orbits. The green points 
                 represent the discovery coordinates of the 13 known objects currently moving in Arjuna-type orbits (Table \ref{discovery}).
                 Only MOIDs (perigees) under 0.05 AU are shown. In this figure, the value of the MOID in AU is colour coded following the 
                 scale printed on the associated colour box.
                }
      \label{known}
    \end{figure}

    Figure \ref{radec} shows that, under the assumptions made here, the best areas to search for these objects are located around right 
    ascension 6$^{\rm h}$ (optimal) and 18$^{\rm h}$, and declination $\in$(0, 30)\degr; the (slightly) best observing epochs are March--May 
    and August--October (see Fig. \ref{prob} in Sect. 7). This is consistent with data in Table \ref{discovery}. There, 9 out of 13 objects
    (69\%) have been found during the favoured epochs. However, and out of 13 known objects (see Table \ref{discovery}) moving within the 
    Arjuna region of the orbital parameter space as defined in Sect. 2, virtually none of them were discovered close to the most favourable 
    search areas in right ascension (see Fig. \ref{known}). The distribution in ecliptic coordinates (see Fig. \ref{lonlat}) is more 
    uniform. The first object was found in 2003 (2003~YN$_{107}$). Nearly 50\% have absolute magnitude $>$ 28 or size smaller than about 
    10~m. 

    A lower limit for the current population size of this dynamical class can be estimated taking into account that three relatively large 
    objects ($H <$ 26 mag) have been discovered in four years within 45\degr of the anti-Sun point (where detection is more efficient, see
    Table \ref{discovery}) and assuming that: (i) all the objects have synodic periods of nearly 43 yr and (ii) that 70\% is the detection 
    effectiveness for ground-based surveys in that area of the sky (remember that the fraction of objects reaching perigee with declination 
    below -30\degr or above +80\degr is $\sim$30\% and it must be discounted). Therefore, we have observed for nearly 10\% of the assumed 
    synodic period with a 70\% efficiency; for that region of the sky, we have already found 7\% of all the brightest objects. In other 
    words, there may be 43 relatively bright Arjunas reaching perigee within 45\degr of the anti-Sun point. But according to our Monte Carlo 
    calculations only 25\% of the objects reach perigee at such large solar elongations. This gives approximately 172 objects with sizes 
    larger than about 30~m currently engaged in this peculiar recurrent resonant behaviour. This number, which is consistent with 
    independent estimates by Brasser \& Wiegert (2008) on the population of asteroids moving in Earth-like orbits, is probably a very 
    conservative lower limit because we do not know exactly how good our current detection rate is nor the actual distribution of periods, 
    sizes, or albedos for these objects. The most critical parameter is the period. Unfortunately, those with very long synodic periods and 
    solar elongations at perigee $<$ 20\degr are the most (potentially) dangerous by far, because they are more difficult to detect. 

    Three very recent discoveries, 2013~BS$_{45}$, 2013~RZ$_{53}$, and 2014~EK$_{24}$, have very long synodic periods (see Table 
    \ref{discovery}). The dynamics of 2013~BS$_{45}$ has been studied by Adamo 
    (2013)\footnote{\scriptsize http://www.aiaahouston.org/Horizons/Horizons\_2013\_03\_and\_04.pdf} and de la Fuente Marcos \& de la 
    Fuente Marcos (2013a). Tiny Aten asteroid 2013~RZ$_{53}$ passed between our planet and the Moon on 2013 September 18. It was first 
    spotted on the 13th at $V$ = 20.2 mag by the Mt. Lemmon Survey (Kowalski et al. 2013). This 1--3 m object ($H$ = 31.2 mag) is unlikely 
    to be of artificial origin and with $a$ = 0.999722 AU, $e$ = 0.048260, and $i$ = 1\fdg5065, it qualifies as member of the Arjuna group 
    as defined above. Asteroid 2014~EK$_{24}$ is a relatively large Apollo ($H$ = 23.2 mag or 60--150 m) discovered on 2014 March 10 at $V$ 
    = 17.9 mag by the Catalina Sky Survey (Larson et al. 2014). So far, 2014~EK$_{24}$ is the largest member of the Arjuna-class. It has $a$ 
    = 1.003690 AU, $e$ = 0.0721495, and $i$ = 4\fdg72541; it is currently engaged in a Kozai resonance with $\omega$ librating around 
    0\degr. Having found three objects with long synodic periods by chance in just over one year hints at the existence of a very large 
    number of similar objects. The long-synodic-period Arjunas may be very numerous. This is consistent with results obtained by Ito \& 
    Malhotra (2010) and Mainzer et al. (2012).

    If most large Arjunas have synodic periods of nearly 100 yr, then their population could be $\sim10^3$ but if the most probable synodic 
    periods are near 1000 yr then the size of the population could be close to 10$^4$. Arjunas can only experience close encounters with 
    the Earth--Moon system. In absence of important non-gravitational and/or secular perturbations, their orbital elements can only be 
    significantly altered during these flybys that are widely spaced in time (at least 43 yr). Arjunas with very long synodic periods can be 
    inherently very stable, dynamically speaking, because the time interval between close encounters could be in the thousands of years or 
    longer. These close approaches are mostly resonant returns (Valsecchi et al. 2003), though. It is, therefore, quite possible that the 
    synodic period distribution for Arjunas is biased in favour of objects with very long periods as they are intrinsically more stable. The 
    extreme case would be represented by those objects trapped at the Lagrangian point L$_3$, at 180\degr ahead of the Earth on its orbit, 
    moving with very low libration amplitudes. These objects, if they do exist, would be stable for very long periods of time and completely 
    harmless for our planet as they cannot experience any close encounters. Objects in 2012~FC$_{71}$-like orbits (de la Fuente Marcos \& de 
    la Fuente Marcos 2013a) are also inherently stable.
%
%
%
%

 \section{Finding them: visibility analysis for Gaia}    
    Gaia (Mignard et al. 2007) is a 5-year mission designed and operated by the European Space Agency (ESA). Since 2014 January 8 it follows 
    a Lissajous orbit around the Earth-Sun Lagrange-point L$_{2}$, in the direction opposite the Sun. The orbit is not shadowed by Earth 
    eclipses. Gaia's primary mission focuses on acquiring astrometry of Galactic stars but it will scan the entire sky, reaching solar 
    elongations as low as 45\degr. Therefore, it will also detect numerous minor bodies, many of them will be new discoveries (Mignard 2002). 
    Gaia's discoveries will have follow-up from the ground. But, how important will likely be its contribution to the study of objects 
    moving in Arjuna-type orbits? 
%
%
     \begin{table*}
      \centering
      \fontsize{8}{11pt}\selectfont
      \tabcolsep 0.1truecm
      \caption{Orbital elements of Gaia on JD 2456800.5 = A.D. 2014-May-23 00:00:00.0 UT (Source: JPL HORIZONS system).}
      \begin{tabular}{ccccc}
       \hline
          $a$ (AU)          & $e$                 & $i$ (\degr)          & $\Omega$ (\degr)   & $\omega$ (\degr)  \\
       \hline
          1.039784463237432 & 0.02092179201154782 & 0.1121955509574652   &  40.55988715506452 & 163.3336670512748 \\
       \hline
      \end{tabular}
      \label{Gaia}
     \end{table*}
%
%
%
%
    \begin{figure}
        \resizebox{\hsize}{!}{
            \includegraphics{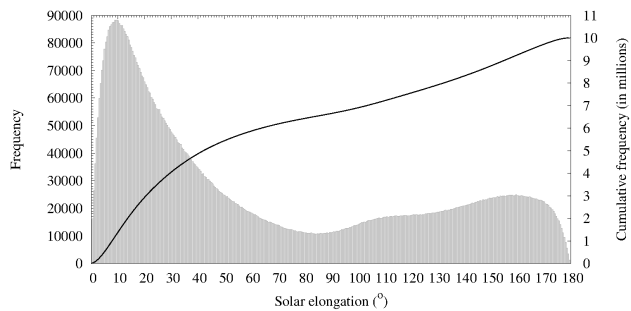}
        }
        \caption{A frequency analysis of the Gaia-centric solar elongation, the bin size is 0\fdg5. Only peri-Gaias $<$ 0.2 AU are 
                 considered. Given the size of the samples studied (10 million test orbits) and the values of the frequencies, error bars 
                 are too small to be seen.
                }
      \label{elongationg}
    \end{figure}
%
%

    If we apply the same Monte Carlo technique used to evaluate the visibility of asteroids moving in Arjuna-type orbits from the ground to 
    the Gaia mission using the available orbit (see Table \ref{Gaia}) for the spacecraft at Epoch JD 2456800.5 (2014 May 23 0:00 UT) we 
    obtain Fig. \ref{elongationg}. Here, we show a frequency analysis of the Gaia-centric solar elongation for objects reaching Gaia-centric 
    distances $<$ 0.2 AU. Nearly 48.5\% of the studied orbits (10$^7$) are viable targets for Gaia if it can reach solar elongations as low 
    as 45\degr. As pointed out above, in theory about 56.4\% of Arjunas with MOIDs under 0.05 AU can be observed from the ground although 
    the actual maximum efficiency is close to 41.4\% (a more realistic estimate is close to 35\%, see above) due to a smaller number of 
    telescopes being located in the southern hemisphere. Therefore, the expected Gaia discovery rate for this class of objects will be, in
    theory, better than that from the ground. 

    Hypothetical Arjunas (and any other Solar System object) will appear as moving objects on Gaia frames. Due to the scanning nature of 
    Gaia's mission (the cadence of the observations), it is unlikely that minor planets will be portrayed in correlated sequences of frames. 
    This will make it difficult for Gaia to independently discover asteroids, particularly at small solar elongations where self-follow-up 
    observations are essential, since ground-based telescopes are unable to observe that region. However, the same object may appear on 
    multiple frames separated by arbitrary time intervals. The analysis of the positions of these apparently random moving points, 
    attempting to fit a Keplerian path to some of them using automated algorithms that link separate observations, may eventually produce a 
    number of high quality orbits over the expected duration of the mission. Therefore, and unfortunately, Gaia is not going to improve 
    dramatically the current discovery rate for minor bodies moving in Arjuna-type orbits and it may not be able to confirm (or disprove) 
    the analysis presented above. However, Gaia will significantly improve our knowledge of Atens and IEOs (Mignard 2002).

 \section{Encountering the Earth at low velocity}
    Close encounters between asteroids moving in Arjuna-type orbits and our planet are characterized by low relative velocities. The rate at 
    which our planet is being hit by minor bodies moving in Arjuna-type orbits depends on the encounter velocities of the objects. During 
    flybys, their paths are deflected towards our planet by its gravitational field. This gravitational focusing effect selectively 
    intensifies low-velocity meteoroid fluxes relative to high-velocity ones (e.g. Lewis 1996); i.e., during close approaches, the Earth's 
    impact cross-section becomes gravitationally enhanced. 

    In order to compute the strength of this enhancement and following \"Opik (1951), we study the gravitationally-enhanced impact 
    cross-section given by the expression $\sigma_{\rm g} = \sigma_{\rm E} \ (1 + v^2_{\rm esc} / v^2_{\rm enc})$, where $\sigma_{\rm E}$ is 
    the geometric cross-section of the Earth, $v_{\rm esc}$ is the escape velocity at an altitude of 100 km above the surface of the Earth 
    (11.1 km s$^{-1}$) and $v_{\rm enc}$ is the unaccelerated effective encounter velocity given by the expression $v^2_{\rm enc} = 
    v^2_{\rm rel} - v^2_{\rm ESC}$, with $v_{\rm rel}$ being the relative velocity at the MOID (perigee) and $v_{\rm ESC}$ is the two-body 
    escape velocity from the MOID. The evaluation of $\sigma_{\rm g}$ requires a statistically significant sample of orbits to obtain the 
    MOIDs, and their associated $v_{\rm ESC}$ and $v_{\rm rel}$. Using the procedure outlined in Sect. 3, $v_{\rm ESC}$ and $v_{\rm rel}$ 
    can then be computed. Although this approach is computer intensive, it produces statistically robust results with very few, if any, a 
    priori assumptions (see above).  
%
%
    \begin{figure}
        \resizebox{\hsize}{!}{ 
            \includegraphics{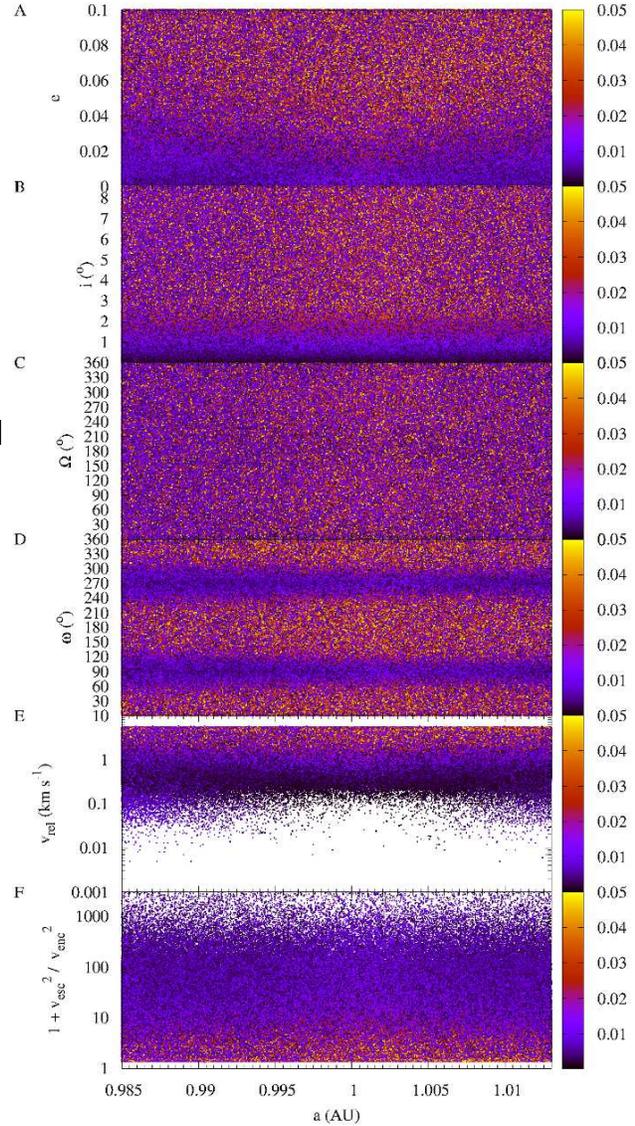}
        }
        \caption{MOID as a function of the various orbital elements of the orbit of the incoming object: semi-major axis, $a$, eccentricity, 
                 $e$, inclination, $i$, longitude of the ascending node, $\Omega$, and argument of perihelion, $\omega$. The value of the 
                 MOID as a function of the relative velocity, $v_{\rm rel}$, of the body moving in an Arjuna-type orbit and the 
                 gravitational focusing factor at the time of the close encounter are also shown. The value of the MOID in AU is colour 
                 coded following the scale printed on the associated colour box. 
                }
      \label{map}
    \end{figure}
%
%
%
%
    \begin{figure}
        \resizebox{\hsize}{!}{ 
            \includegraphics{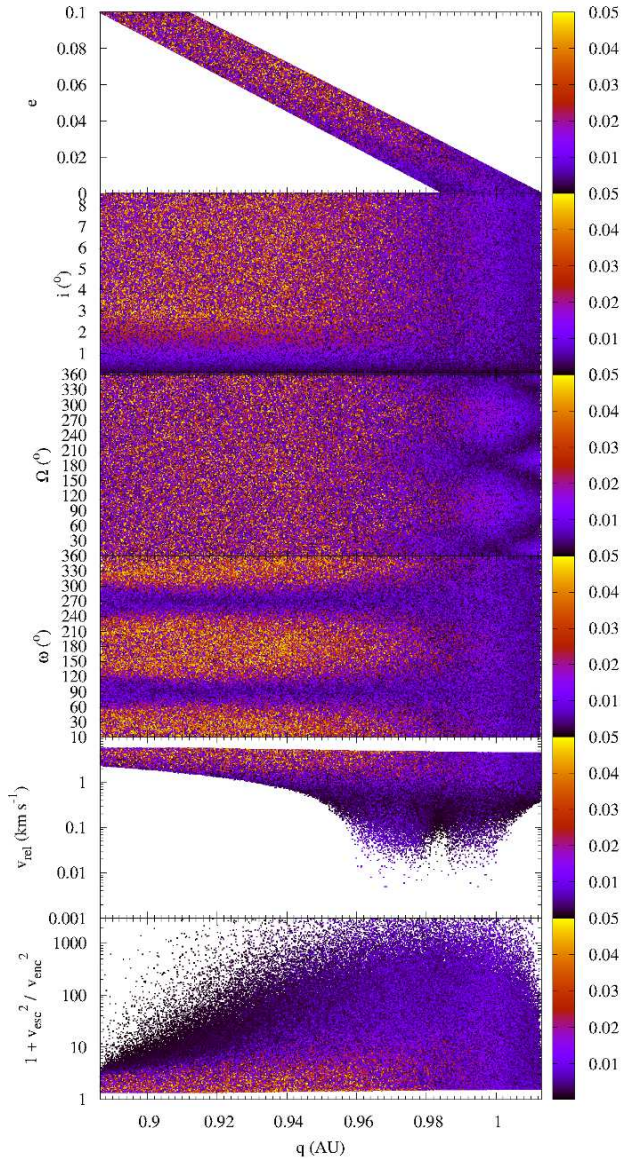}
        }
        \caption{As Fig. \ref{map} but in terms of the perihelion distance, $q$. 
                }
      \label{mapq}
    \end{figure}
%
%

    Figure \ref{map} summarizes our results for the gravitationally-enhanced impact cross-section (the gravitational focusing factor, 
    $\sigma_{\rm g}/\sigma_{\rm E}$) of objects moving in Arjuna-type orbits as a function of the distance of closest approach (panel F). 
    Only MOIDs under 0.05 AU are shown and this represents over 90\% of the orbits studied (10$^{7}$). In this figure, the value of the MOID 
    in AU is colour coded following the scale printed on the associated colour box. In addition, the orbital elements of the incoming object 
    are also plotted as a function of the MOID. As usual, the orbital elements follow a uniform distribution within the ranges indicated in 
    Sect. 3. As expected, the MOIDs are smaller for orbits with values of the argument of perihelion similar to Earth's perihelion 
    ($\sim$258\degr) and aphelion ($\sim$78\degr), see panel D. The relative velocities, $v_{\rm rel}$, of the bodies moving in Arjuna-type 
    orbits (panel E) at perigee can be as low as a few m~s$^{-1}$. Due to these extremely low encounter velocities, the gravitational 
    focusing factor reaches values higher than 1,000 (0.053\%). Over 6.7\% of objects experience gravitational focusing factors in the range 
    10--100. This is consistent with results obtained by Wetherill \& Cox (1985) and Kortenkamp (2013). Over 7.9\% of the orbits have 
    $v^2_{\rm rel} < v^2_{\rm esc}$ at perigee and we interpret this number as the overall fraction of Arjunas that can be captured as 
    transient natural satellites. The existence of short-lived natural satellites of the Earth was first discussed by Baker (1958) but 
    actual objects were found only recently (Kwiatkowski et al. 2009; Granvik et al. 2012). Plans to perform a systematic search for these 
    so called minimoons using the Arecibo and Greenbank radio telescopes are under way (Bolin et al. 2014).

    If we compare the distribution of perigees in Fig. \ref{map} as a function of the various orbital elements with those of actual objects 
    in Table \ref{members}, we realize that the vast majority of objects with the lowest perigees are expected to have eccentricity under
    0.02, inclination under 2\degr, and argument of perihelion close to 90\degr or 270\degr. Data in Table \ref{members} clearly show that
    the current sample of Arjunas is biased in favour of objects with low inclinations and argument of perihelion close to 90\degr or 
    270\degr. Both biases are consistent with the expectations because objects moving in orbits characterized by shorter perigees are easier
    to identify. However, Fig. \ref{map} also predicts an excess of low-eccentricity orbits among those of objects with the shortest 
    perigees and this is not observed. For real data, an excess of minor bodies in comparatively eccentric orbits is found instead. Given 
    the fact that our analysis assumes uniform distributions for the orbital elements, it is obvious that the observed excess of objects in 
    relatively eccentric orbits may represent a distinctive feature of this population. Eccentricities can be excited as a result of secular 
    resonances. However, the surprising excess of objects with eccentricities $>$ 0.08 could also be connected with the dominant supply 
    mechanism for objects in this group. Debiased samples--see Mainzer et al. (2011), Greenstreet et al. (2012), or Greenstreet \& Gladman 
    (2013)---show more objects in more eccentric orbits but also more inclined. In principle, additional observational data are needed to 
    explain this excess. 

\begin{figure}
        \resizebox{\hsize}{!}{
            \includegraphics{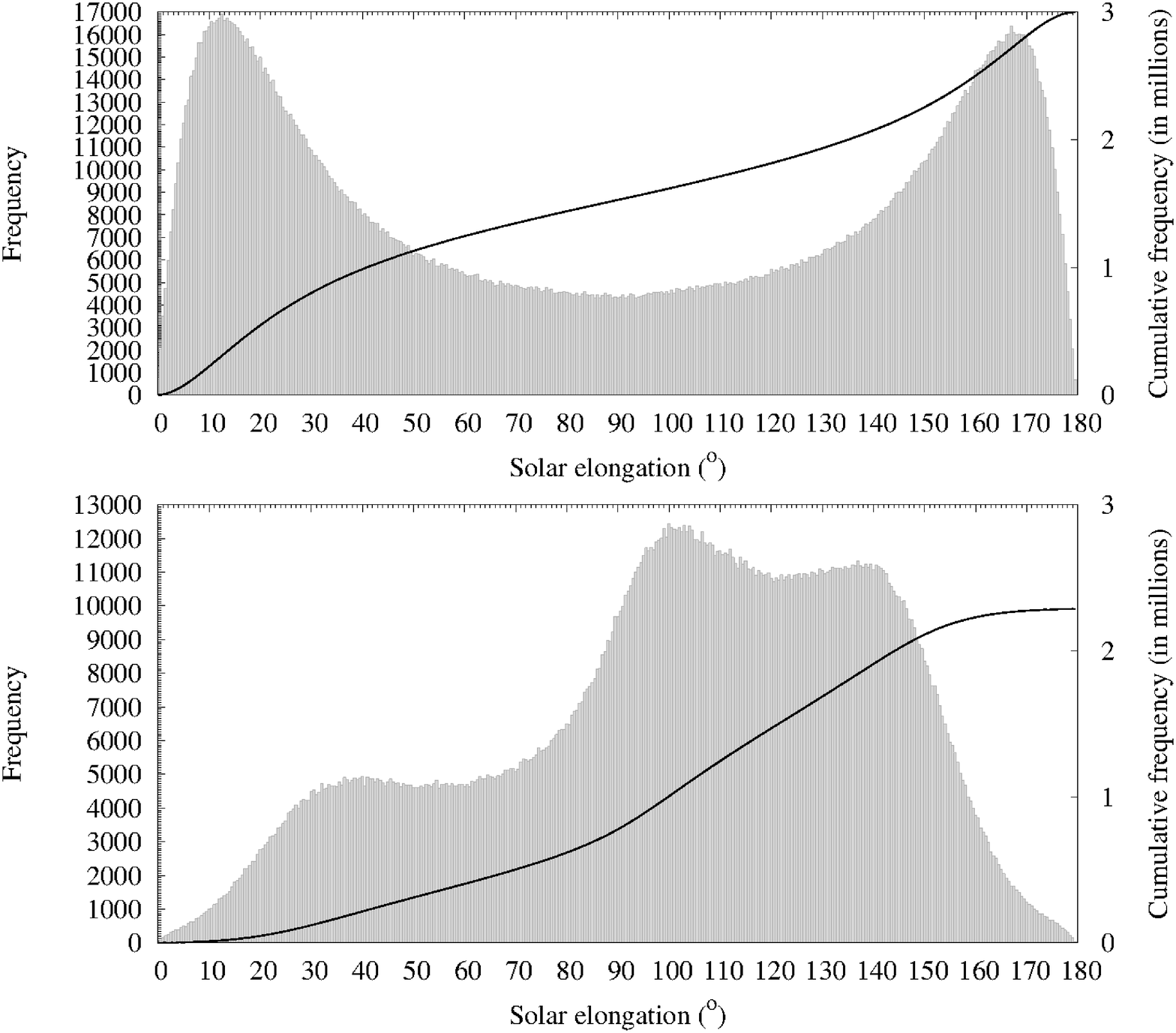}
        }
        \caption{Frequency distribution in solar elongation at perigee of Arjuna-type orbits for $e < 0.03$ (top panel) and $e > 0.07$
                 (bottom panel). The bin size is 0\fdg5. Only perigees $<$ 0.05 AU are considered. Given the size of the samples studied and 
                 the values of the frequencies, error bars are too small to be seen.
                }
      \label{eloe}
    \end{figure}

\begin{figure}
        \resizebox{\hsize}{!}{
            \includegraphics{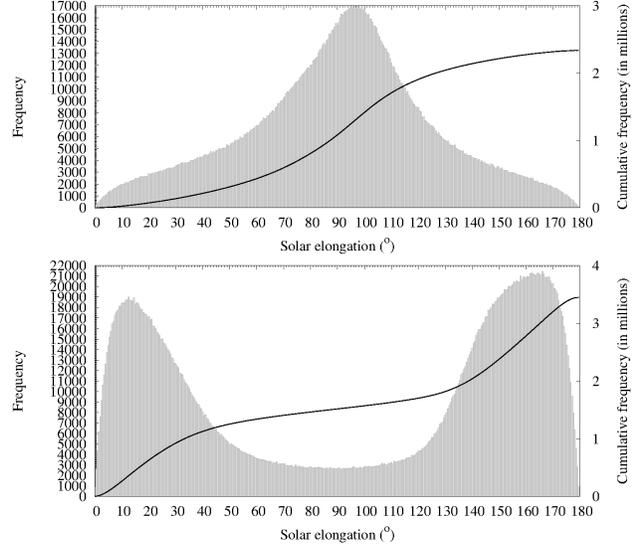}
        }
        \caption{Frequency distribution in solar elongation at perigee of Arjuna-type orbits for $i <$ 2\degr (top panel) and $i >$ 5\degr
                 (bottom panel). The bin size is 0\fdg5. Only perigees $<$ 0.05 AU are considered. Given the size of the samples studied and 
                 the values of the frequencies, error bars are too small to be seen.
                }
      \label{eloi}
    \end{figure}

    However, we have already found that the detectability of Arjunas depends strongly on the value of their solar elongation at perigee (see 
    Fig. \ref{elongation}) but we did not study how this parameter depends on orbital elements like eccentricity or inclination. If we 
    repeat the frequency analysis in Fig. \ref{elongation} for the low and high ends of the distribution in $e$ and $i$ we obtain Figs 
    \ref{eloe} and \ref{eloi}. Figure \ref{eloe} shows the distribution in solar elongation for orbits with $e < 0.03$ (top panel) and $e > 
    0.07$ (bottom panel). It is clear that the excess of eccentric orbits among the observed sample is the result of an observational bias. 
    More eccentric orbits tend to reach perigee at more favourable solar elongations; i.e. the objects following them are easier to 
    discover. As for the inclination, Fig. \ref{eloi} shows the distribution in solar elongation for orbits with $i <$ 2\degr (top panel) 
    and $i >$ 5\degr (bottom panel). Here, the bias in favour of any of the ends of the distribution is not as strong as in the case of the 
    eccentricity. Nearly 40\% of the hypothetical high-inclination Arjunas reach perigee too close to the Sun to be discovered from the 
    ground but also a large fraction of the low-inclination Arjunas reach perigee close to 90\degr from the Sun. Objects moving in orbits 
    near the low end of the eccentricity distribution and the high end of the inclination distribution are markedly more difficult to 
    identify because a large fraction of them reach perigee within 50\degr from the Sun. These objects are absent from the observed sample 
    simply because they are intrinsically difficult to discover when observing from the ground (or Gaia). Also, they cannot be observed 
    before they strike. 
%
%
%
%

    Following Gronchi \& Valsecchi (2013a), Fig. \ref{mapq} is a replot of Fig. \ref{map} in terms of the perihelion distance. The regions 
    with the shortest MOIDs follow patterns similar to those displayed in Figs 1, 17, 18, 19 and 22 in Gronchi \& Valsecchi (2013a) by 
    near-Earth asteroids with absolute magnitude $H >$ 22. Particularly interesting is the panel associated to $\Omega$ that explains what 
    is observed in Fig. \ref{histo} as the result of an observational bias, objects with the shortest perigees are the easiest to detect.
%
%
%
%

    Regarding Fig. \ref{map}, it may be argued that the MOID is not uniformly zero due to poor sampling in the Monte Carlo calculation. This
    is, however, a very simplistic interpretation. It is true that if two Keplerian orbits intersect, it is always possible to find angles 
    $\Omega$, $\omega$ and $f$ so that the minimal approach distance is zero. But just a fraction of our simulated Arjuna-type orbits 
    actually intersect the orbit of our planet, those in the Apollo and Aten classes; the orbits of objects following Amor- or IEO-type 
    trajectories do not intersect Earth's and, therefore, their minimal approach distances (even if small) are always greater than zero.  

 \section{Impact cross-section: a comparison}
    By definition, potentially hazardous asteroids have MOIDs with the Earth of 0.05~AU or less (and their absolute magnitude, $H$, must be 
    equal to 22 or brighter); for the group of objects discussed here, the (geometrical) probability of having a MOID that small is over 
    90\%. Arjunas in Tables \ref{members} and \ref{discovery}, with the exception of 2012~FC$_{71}$, have MOIDs under 0.05~AU but they all 
    have $H > 22$ mag. Therefore, they mostly move in PHA-like orbits but they do not qualify as official PHAs because they are too small; 
    they represent the PHA regime. Our Monte Carlo calculations show that the encounter velocities are in some cases so low that the 
    gravitational focusing factor could be over 3,000 ($<$ 0.016\%). This clearly indicates that compared to typical Apollo or Aten 
    asteroids, the probability of collision for Arjuna-type orbits is much higher, but how much higher? 

\begin{figure}
        \resizebox{\hsize}{!}{
            \includegraphics{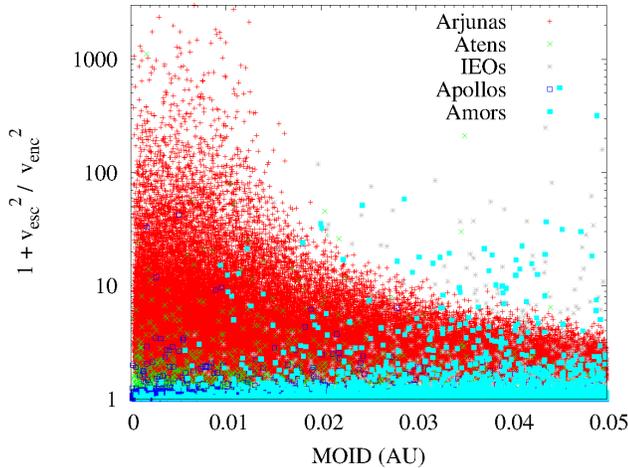}
        }
        \caption{Gravitational focusing factor for Arjunas, Amors, Apollos, Atens and IEOs.
                } 
      \label{sigma}
    \end{figure}

\begin{figure}
        \resizebox{\hsize}{!}{
            \includegraphics{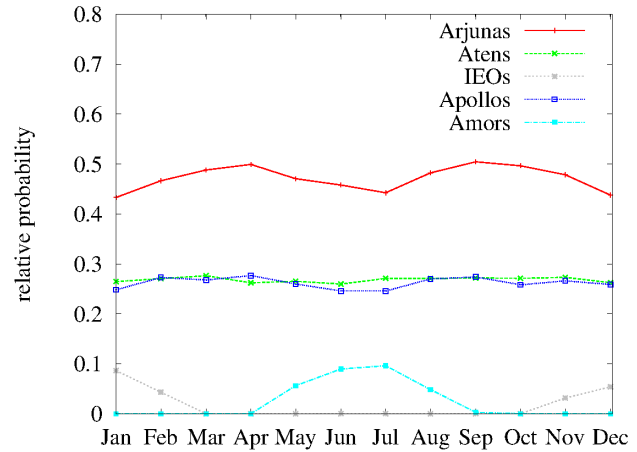}
        }
        \caption{Monthly relative probability of having a close encounter with MOID $<$ 0.0098 AU for Arjunas, Amors, Apollos, 
                 Atens and IEOs. The relative probability is the number of MOIDs $<$ 0.0098 AU divided by the number of MOIDs $<$ 0.05 AU,
                 which is the threshold for the PHA regime. Given the large size of the simulated samples, errors bars are as large as the
                 size of the actual points.
                }
      \label{prob}
    \end{figure}

    In order to answer the question posed above in a relevant and meaningful fashion, we consider the following asteroid populations and 
    parameter spaces, where $q = a (1 - e)$ is the perihelion distance and $Q = a (1 + e)$ is the aphelion distance: Amors ($a >$ 1 AU, 
    1.017 $< q$(AU)$<$ 1.3), Apollos ($a >$ 1 AU, $q <$ 1.017 AU), Atens ($a <$ 1 AU, $Q >$ 0.983 AU) and IEOs ($a <$ 1 AU, 0.72 $< Q$(AU)
    $<$ 0.983). We also use the same Monte Carlo approach applied to Arjunas. The fraction of objects with MOID under 0.05~AU for Atens is 
    29.6\%, for IEOs is 3.4\%, for Apollos is 24.9\%, and for Amors is 3.5\%. Figure \ref{sigma} shows the gravitational focusing factor for 
    these populations. For non-Arjunas, it is rarely $>$ 10 ($<$ 0.05\%); for Arjunas, almost 7\% of the orbits experience focusing factors 
    that high or higher. Therefore, the fraction of Arjunas having their impact cross-sections significantly enhanced is over two orders of 
    magnitude larger than that of the general NEO population which is fully consistent with results obtained by Brasser \& Wiegert (2008) 
    using actual integrations. Figure \ref{prob} displays the relative probability of having a close encounter with MOID $<$ 0.0098 AU (one 
    Hill radius) for objects with MOIDs under 0.05 AU (the threshold for the PHA regime) as a function of the time of the year. Seasonal 
    variations are obvious. As expected, IEOs can encounter the Earth more easily when our planet is near perihelion and Amors when the 
    Earth is at aphelion.

    The mean impact probabilities associated to Amors, Apollos, and Atens have been studied previously (see e.g. Steel \& Baggaley 1985; 
    Chyba 1993; Steel 1995; Dvorak \& Pilat-Lohinger 1999; Dvorak \& Freistetter 2001) and it is well established that Atens have the 
    shortest lifetimes against collision, closely followed by Apollos (a factor 2--5 longer) and then Amors, three orders of magnitude 
    longer. Some of these estimates are based on actual statistics of observed objects so they are biased against fainter minor bodies. Our 
    Monte Carlo results assuming uniformly distributed orbital elements suggest similar probabilities for Apollos and Atens, and 
    significantly lower probabilities for Amors and IEOs with large seasonal variations for these two. Arjunas have higher probabilities, in 
    theory, but their encounter velocities are, in some cases, so low, that captures as transient natural satellites are probably more 
    likely than actual impacts: the probability of capture is nearly 0.08 but the probability of collision is 1.2$\times$10$^{-6}$. These 
    numbers are for unperturbed orbits, i.e. they are geometrical probabilities. They have been obtained in the usual way, dividing the 
    number of relevant events by the total number of orbits studied. 
%
%
%
%
%
%
%
%

 \section{Do the Arjunas dominate the overall flux of Earth's impactors at sizes under 50~m?}
    Our planet accretes thousands of tonnes of microscopic dust per year (Whipple 1950; Love \& Brownlee 1993). The results of simulations 
    of the dynamical evolution of interplanetary dust grains near the orbit of the Earth completed by Dermott et al. (1994), Kortenkamp \& 
    Dermott (1998a,b) and Kortenkamp (2013), indicate that a significant fraction of the material accreted each year experienced previous 
    trapping in a 1:1 mean motion resonance with our planet. This material is moving in low-eccentricity, low-inclination orbits, typical of 
    orbitally decaying dust particles from the asteroid belt (Kortenkamp \& Dermott 1998a). The accretion rate for this resonant material 
    is 10--3000 times larger than that of dust from other sources (Kortenkamp 2013). If dust grains trapped in a 1:1 mean motion resonance 
    are the current primary source of accreted material for the Earth, can larger objects moving in similar orbits be the main source of 
    Earth impactors? Is it possible that asteroids moving in Arjuna-type orbits may dominate the overall flux of Earth's impactors at small 
    sizes?

    Most authors would say that there is no evidence that Arjunas may dominate the flux of Earth impactors, even in the 10--50-m range. The 
    impact frequency does not depend solely on the encounter velocity (and hence on the gravitationally-enhanced collisional cross-section); 
    it also depends on the number of objects in the population and their steady state orbital distribution. Analyses based on Bottke et 
    al.'s (2002) NEO model show that Arjunas are a very small sub-population of near-Earth asteroids. 

    Chesley \& Spahr (2004) found that the steady-state sky-plane distribution of Earth-impacting asteroids long before impact is 
    concentrated on the ecliptic and towards small solar elongations. They performed an extensive study of the impactor population but did 
    not focus on Earth-like orbits. Their study found that the hazard fraction for Earth-like orbits is the highest but they represent a 
    very small percentage of the impacting population. Vere\v{s} et al. (2009) have studied the detection efficiency of future surveys aimed 
    at identifying Earth-impacting asteroids. They show that theoretical models and actual observations of sporadic background fireballs do 
    not favour an excess of impactors moving in Arjuna-type orbits. Sporadic background fireballs tend to have semi-major axes in the range 
    2--3 AU, eccentricities in the ranges 0.6--0.8 and 0.9--1.0, and inclinations below 20\degr.

    It is true that of all the meteorites for which the orbits have been determined thanks to the images of camera networks none has an 
    Arjuna-like orbit but it is also true that camera networks operate at night and that our calculations (see Sects. 3 and 4) show that 
    nearly 50\% of the objects moving in Arjuna-type orbits reach perigee at solar elongations $<$ 90\degr. Campbell-Brown (2008) has 
    published results from the analysis of the orbits of sporadic meteoroids detected using radar. These observations are unbiased with 
    respect to the value of the solar elongation of the meteoroid but they only provide information on the orbits of very small objects that 
    actually collided with our planet not casual flybys by metric-sized Arjunas. Nonetheless, the results from that study are worth 
    discussing here. Figure 10 in Campbell-Brown (2008) shows a peak at low-inclination, semi-major axis $\sim$1 AU orbits. However, the 
    eccentricity distribution has a maximum at 0.7. In principle, these results (lack of meteors moving in low-eccentricity orbits) argue 
    against the recent numerical work of Kortenkamp (2013) but observational biases are also at work for radar data. If the orbit of a 
    meteoroid is only slightly larger than that of the Earth, as it is the case for nearly 50\% of all hypothetical meteoroids moving in 
    Arjuna-type orbits, the encounter speed in a collision with the back (from the west, antapex) of the Earth will be low. Slower meteors 
    produce less ionization and, therefore, are much less likely to be observed (using either radar or optical equipment) than higher 
    encounter speed meteors moving in very eccentric orbits.  

    Observations are clearly biased against meteors that encounter the Earth moving in Arjuna-like orbits, in particular the 
    low-eccentricity feature is very problematic. Therefore, the lack of identified Arjuna meteors cannot be used to argue against an 
    eventual excess of Arjunas in the overall flux of Earth's impactors. In fact, Arjuna-like meteoroids have been identified. Using 
    Arecibo's antenna, Janches et al. (2000) found that the fraction of near-antapex micrometeors moving slower than the Earth escape 
    velocity is nearly 39\%; they identified one Arjuna-like micrometeor (no. 1 in their Table II) among the objects moving faster than the 
    escape velocity. This means that nearly 44\% of detections in their study were following Arjuna-like orbits. In any case, the available
    evidence is far from conclusive. Our own geometrical analysis (see above) indicates that, among Arjunas, capture as transient minimoons
    is far more probable than having an impact on our planet.  

 \section{Discussion}
    One of the most puzzling facts in Table \ref{discovery} is the lack of relatively bright Arjunas. None of the known objects moving in 
    Arjuna-type orbits is brighter than $H$ = 23 mag. The current number of Earth co-orbital NEOs has been estimated by Morais \& 
    Morbidelli (2002, 2006). The model in Morais \& Morbidelli's papers is based on a previous model by Bottke et al. (2000, 2002) which 
    only applies in the size range corresponding to $H < 22$. This limit (about 400 m in diameter) reflects the transition from tensile 
    strength dominated bodies to rubble piles or gravity-dominated aggregates (see, e.g., Love \& Ahrens 1996). Morais \& Morbidelli (2002, 
    2006) predict 16.3$\pm$3.0 objects with $H < 22$ currently engaged in co-orbital motion with the Earth. The fraction of bodies with
    eccentricity $< 0.1$ is $<$ 0.02 and the fraction with inclination $<$ 10\degr is $<$ 0.05 (Morais \& Morbidelli 2002). According to 
    their results no Earth co-orbitals with $H < 22$, $e < 0.1$, and $i <$ 10\degr should be observed, and this is exactly what we have in 
    Table \ref{discovery}. 

    Even if data in Table \ref{discovery} are affected by observational biases (see above) and the number of known objects is small, it 
    could still be useful to have some estimate of the expected number of smaller objects moving in dynamically cold orbits as a function of 
    $H$. Following Morais \& Morbidelli (2006), we assume a power law of the form $N(H) \propto 10^{\alpha H}$ and use data in Table 
    \ref{discovery} to find the constant and the index $\alpha$. We obtain
    \begin{equation}
        N(H < H_o) = 1.6\times10^{-7} \ 10^{0.284\ H} \,, \label{law}
    \end{equation}   
    where $N(H < H_o)$ is the number of objects with absolute magnitude $< H_o$. The correlation coefficient is $r$ = 0.9987. This simple 
    law predicts seven objects with $H < 27$ (where seven are observed), 53 with $H < 30$ and more than 10$^{3}$ with $H < 35$. These 
    numbers must be multiplied by a factor two, at least, to take into account the observational bias associated to the solar elongation. We 
    do not attempt a correction to account for the distribution in synodic period. On the other hand, our model is of the form $\log N(H) = 
    \log K + \alpha H$, with $\log K$ = -6.8$\pm$0.3 and $\alpha$ = 0.284$\pm$0.010 mag$^{-1}$. Therefore, the dispersions are relatively 
    large and, in consequence, the predicted sizes of the Arjuna population as a function of the absolute magnitude are affected by large 
    errors (for example, $N(H < 27)$ is in the range 2--27). In any case, the available observational evidence suggests that the number of 
    metric-sized Arjunas is, at least, 10$^3$ and probably larger.
     
    Any object currently moving in Arjuna-type orbits must be dynamically young because this population is rapidly depleted as a result of 
    their high impact probability with the Earth and intrinsically chaotic dynamics (Koon et al. 2001). In fact, the mere existence of 
    present-day objects in these inherently unstable orbits clearly requires one or more mechanisms able to replenish the continuous losses. 
    The supply mechanisms must be efficient because 13 members of this group have already been discovered and several others may be recent 
    ejections from the Arjuna region of the orbital parameter space (de la Fuente Marcos \& de la Fuente Marcos 2013a). Their small sizes 
    (see Table \ref{discovery}) suggest that many of them could be fragments of fragments. 

    Asteroidal decay could be induced by collisional processes (e.g. Dorschner 1974; Ryan 2000) but also be the combined result of thermal 
    fatigue (e.g. \v{C}apek \& Vokrouhlick\'y 2010) and rotational (e.g. Walsh, Richardson \& Michel 2008) or tidal stresses (e.g. T\'oth, 
    Vere\v{s} \& Korno\v{s} 2011). These last three processes can easily produce secondary fragments. Planetary ejecta from Mars, the 
    Earth--Moon system, and Venus are also a possible source (Warren 1994; Gladman et al. 1995; Bottke et al. 1996; Gladman 1996, 1997) as 
    the Kozai mechanism (Kozai 1962) may be able to stabilise some orbits (2012~FC$_{71}$-like, de la Fuente Marcos \& de la Fuente Marcos 
    2013a). The role of the Kozai effect arising from Jovian perturbations on the NEO population has also been discussed in JeongAhn \& 
    Malhotra (2014). On the other hand, Mainzer et al. (2014) has reported physical parameters for tiny NEOs, some of them as small as 8~m 
    in diameter. The sample appears to show an increase in albedo with decreasing diameter. Although this could be explained as the result 
    of an observational bias against small and dark NEOs, it may also indicate that smaller objects have fresh and bright surfaces because 
    they are the result of recent fragmentation caused by any of the mechanisms pointed out above.

    But, if the impact risk associated to these objects is so large, how is it possible that no impacts from this group have ever been 
    recorded? The actual impact frequency depends on the gravitationally-enhanced collisional cross-section but also on the number of 
    objects in the population and the distribution of their orbital elements. Our analysis is purely geometrical, we do not know how many 
    objects are members of this population nor how their orbital distributions are. Our study demonstrates that the presently known 
    population is heavily biased in terms of eccentricity, inclination, longitude of the ascending node, and argument of perihelion. We also 
    know that Arjuna meteors entering the atmosphere are intrinsically difficult to identify because they tend to be underluminous and 
    nearly half of them will find our planet during daytime. However, and even if it is true that they have not made headlines recently, 
    Arjuna impacts have happened in the past and they will happen again. 

    One key event to realise this fact is in the widely observed, long duration train of meteors of 1913 February 9 (Chant 1913a,b). This 
    multiple fireball episode was originally interpreted by Chant as the result of the fall of a body that was orbiting the Earth. A 
    temporary natural satellite was also suggested as the source of the meteor procession by O'Keefe (1959, 1960) and Bagby (1966), among 
    others. The only group of asteroids known to have a non-negligible chance of becoming ephemeral Earth satellites are the Arjunas. 
    Another piece of evidence is in the actual record of observed meteorite falls. Dodd \& Lipschutz (1993) and Lipschutz, Wolf \& Dodd 
    (1997) have found clustering among CM chondrite meteorites fallen between 1921 and 1969. They have explained the unusual distribution of 
    meteorite falls as resulting from a stream (one or more) of meteoroids moving in nearly circular, Earth-like paths. That can be 
    interpreted as a signature of the presence of coherent groups of asteroidal fragments moving in Arjuna-type orbits. 
     
    The analysis presented here is mostly assessing the geometrical risk of Arjuna asteroid impacts but the real risk is unknown and it may 
    be higher than the one estimated here because we do not know the actual size of this unusual asteroid population. If the size of this 
    population is small, as most studies predict, then the risk is completely negligible. The lack of answers to fundamental questions 
    prevents a deeper analysis. A number of plausible mechanisms able to supply members to this population have been pointed out above but 
    it is unclear which ones are dominant. The actual shape of the size function (the expected number of objects as a function of the 
    absolute magnitude) is also likely linked to these mechanisms. Understanding where they come from, how many there are today, and how 
    they evolve is vital to fully evaluate how many of them move in Earth-threatening trajectories. We cannot answer these questions based 
    on the properties of just 13 known objects. If we want to better quantify the real risk associated to this population, specific search 
    programs must be implemented.  

 \section{Conclusions}
    There are no objects moving in Arjuna-type orbits among the minor bodies currently listed by NASA as PHAs. This is not because objects
    moving in Arjuna-type orbits cannot experience close approaches to our planet; in fact, their probability of having a MOID with the 
    Earth of 0.05~AU or less is higher than 0.9. However, and in order to be catalogued as PHA, the object's MOID with the Earth must be 
    0.05~AU or less and its $H$ must be 22 mag or brighter. None of the known objects moving in Arjuna-type orbits is brighter than 23 mag. 
    This could be a mere coincidence but it is consistent with current models (Morais \& Morbidelli 2002). However, it may also be the 
    result of the inherently chaotic dynamics characteristic of this population. Periodic, but not frequent due to their long synodic 
    periods, close encounters with the Earth--Moon system or collisions may contribute to the disruption of strengthless rubble piles and 
    the subsequent formation of secondary fragments (see, e.g., Grun et al. 1985; Hahn \& Rettig 1998) that tend to remain within the Arjuna 
    orbital parameter domain for thousands of years.  

    Our numerical study uses Monte Carlo techniques to study statistically the orbital properties of Arjuna asteroids and their implications. 
    The Monte Carlo approach allows extensive coverage of the orbital parameter space without any loss of generality and, more importantly, 
    without making any unwarranted assumptions. The main conclusions of our work can be summarized as follows:
    \begin{itemize}
       \item For ground-based observations, the solar elongation at perigee of nearly half of the Arjunas is less than 90\degr. They are 
             best observed by space-borne telescopes. Gaia is not going to improve significantly the current discovery rate for this class 
             of objects.
       \item The currently known Arjuna population is heavily biased in favour of objects with argument of perihelion close to 90\degr or
             270\degr because these minor bodies have the shortest perigees and, therefore, the smallest apparent magnitudes. Brighter 
             objects are always easier to identify.
       \item The currently known Arjuna population is biased in favour of objects with orbital inclination less than 2\degr because Arjunas 
             with low inclination tend to have shorter perigees and also smaller apparent magnitudes.
       \item The currently known Arjuna population shows an excess of members with eccentricities $>$ 0.08. In general, these objects do not 
             have the shortest perigees but they are overabundant because they reach perigee at favourable solar elongations. 
       \item Arjunas moving in orbits with $e <$ 0.03 and $i >$5\degr are intrinsically difficult to discover due to observational bias. 
             Objects in this group that also have long synodic periods are virtually invisible to current surveys.  
       \item Based on the available observational data and our Monte Carlo analysis, the number of Arjunas with sizes larger than 30 m is at 
             least 172. In our opinion, this is a very conservative estimate as it depends strongly on the distribution of synodic periods. 
       \item Arjunas with very long synodic periods are inherently more dynamically stable than those with periods close to the lower end of 
             the distribution (43 yr). In absence of significant secular perturbations, this may result in an excess of Arjunas with very 
             long synodic periods. If they do indeed exist, the number of Arjunas with sizes larger than 10 m could be in the tens of 
             thousands. 
       \item Arjunas have intrinsically low encounter velocities with the Earth. This induces a 10--1,000-fold increase in their impact 
             cross-section with respect to what is typical for objects in the Apollo or Aten asteroid populations.
       \item The geometric probability of capture of Arjunas as transient natural satellites of our planet is about 8\%.
       \item Arjunas are more likely to become transient satellites than Earth's impactors.
       \item Although, geometrically speaking, the Arjunas are expected to dominate the overall flux of Earth's impactors at sizes under 
             50~m, the available observational evidence---which is biased and far from complete---neither disprove nor confirm this
             geometrical expectation.
    \end{itemize}
    The most intimidating of the unknown objects that may in time collide with our planet are those intrinsically difficult to discover. 
    Besides small long-period comets, asteroids moving in Earth-like orbits have the largest probability of not being detected before impact 
    due to their small sizes and long synodic periods. Long synodic periods seriously reduce our ability to discover and track NEOs. These 
    objects may remain undiscovered despite having intensive minor planet search programmes running for decades. It is the conventional 
    wisdom that any NEO moving in a collision course with the Earth will suffer multiple close encounters with our planet before it hits 
    and, in consequence, it should be eventually discovered by extensive surveys. In contrast, our analysis indicates that the most 
    observationally challenging of the Arjunas are unlikely to be discovered in significant numbers by ordinary, ground-based NEO surveys 
    during the next several decades. It is indeed unfortunate that the most elusive minor bodies are also the ones with the highest 
    theoretical impact probability; luckily, they are all small objects. The few identified so far may represent just the tip of the 
    iceberg. All of them have been discovered from the ground although no specific program aimed at these objects exists yet. The optimal 
    observation strategy for discovering these objects is not currently known but our study provides some leads. We confirm that only 
    space-based surveys can efficiently find and track these targets.
     
 \begin{acknowledgements}
    We would like to thank the referee, G. F. Gronchi, for advise and some useful suggestions.
    This work was partially supported by the Spanish `Comunidad de Madrid' under grant CAM S2009/ESP-1496. 
    We thank M. J. Fern\'andez-Figueroa, M. Rego Fern\'andez, and the Department of Astrophysics of the 
    Universidad Complutense de Madrid (UCM) for providing computing facilities. Most of the calculations 
    and part of the data analysis were completed on the `Servidor Central de C\'alculo' of the UCM and we 
    thank S. Cano Als\'ua for his help during this stage. In preparation of this paper, we made use of the 
    NASA Astrophysics Data System, the ASTRO-PH e-print server and the MPC data server.     
 \end{acknowledgements}

\end{document}